\documentclass[acmsmall, authorversion, printcopyright=false, nonacm]{acmart}

\usepackage[colorinlistoftodos,bordercolor=white,color=green!40,textwidth=2cm]{todonotes}
\usepackage{multirow}
\usepackage{colortbl}
\usepackage{subcaption}
\usepackage{listings}
\usepackage{xcolor}
\usepackage{amsmath}
\usepackage{algorithm}
\usepackage{algorithmic}
\newcommand\algorithmicprocedure{\textbf{Procedure}}
\newcommand{\algorithmicendprocedure}{\algorithmicend\ \algorithmicprocedure}
\makeatletter
\newcommand\PROCEDURE[3][default]{%
  \ALC@it
  \algorithmicprocedure\ \textsc{#2}(#3)%
  \ALC@com{#1}%
  \begin{ALC@prc}%
}
\newcommand\ENDPROCEDURE{%
  \end{ALC@prc}%
  \ifthenelse{\boolean{ALC@noend}}{}{%
    \ALC@it\algorithmicendprocedure
  }%
}
\newenvironment{ALC@prc}{\begin{ALC@g}}{\end{ALC@g}}
\newcommand{\Input}[0]{\hspace*{\algorithmicindent} \textbf{Input}}
\newcommand{\Output}[0]{\hspace*{\algorithmicindent} \textbf{Output}}
\newcommand{\Desc}[2]{\hspace*{\algorithmicindent} \makebox[2em][l]{#1}#2}
\newcommand{\MyComment}[1]{\hfill{{\tt// #1}}}
\newcommand{\MyLineComment}[1]{\STATE{{\tt// #1}}}
\newcommand{\textkw}[1]{\textcolor{blue}{\tt#1}}
\newcommand{\example}[1]{%
    \vspace{0.15cm}%
    \noindent\fcolorbox{blue}{blue!20!white}{\parbox{0.98\textwidth}{%
        \color{black}{%
            \underline{\textbf{\texttt{DummyModule} example}}\hspace{0.6\textwidth} (\textit{cf} Listing \ref{sec.space:list.annotations})\\%
            #1%
            }}}%
            \vspace{0.15cm}
    }
\newcommand{\definition}[1]{#1}
\newcommand{\ie}{{\it i.e. }}
\newcommand{\eg}{{\it e.g. }}
\newcommand{\etal}{{\it et al. }}
\newcommand{\ccg}{\cellcolor{black!30!white}}
\newcommand{\inred}[1]{\textcolor{red}{#1}}

\begin{document}

\title{A Chisel Framework for Flexible Design Space Exploration through a Functional Approach}

\author{Bruno Ferres}
\affiliation{%
  \institution{Univ. Grenoble Alpes, CNRS, Grenoble INP\textsuperscript{*}\authornote{Institute of Engineering Univ. Grenoble Alpes}, TIMA}
  \streetaddress{46 Avenue F\'elix Viallet}
  \city{Grenoble}
  \country{France}
  \postcode{F-38000}
}
\email{bruno.ferres@grenoble-inp.org}
\author{Olivier Muller}
\affiliation{%
  \institution{Univ. Grenoble Alpes, CNRS, Grenoble INP\textsuperscript{*}, TIMA}
  \streetaddress{46 Avenue F\'elix Viallet}
  \city{Grenoble}
  \country{France}
  \postcode{F-38000}
}
\email{olivier.muller@univ-grenoble-alpes.fr}
\author{Fr\'ed\'eric Rousseau}
\affiliation{%
  \institution{Univ. Grenoble Alpes, CNRS, Grenoble INP\textsuperscript{*}, TIMA}
  \streetaddress{46 Avenue F\'elix Viallet}
  \city{Grenoble}
  \country{France}
  \postcode{F-38000}
}
\email{frederic.rousseau@univ-grenoble-alpes.fr}

\begin{abstract}
    As the need for efficient digital circuits is ever growing in the industry, the design of such systems remains daunting, requiring both expertise and time.
    In an attempt to close the gap between software development and hardware design, powerful features such as functional and object-oriented programming have been used to define new languages, known as Hardware Construction Languages.
    In this article, we investigate the usage of such languages --- more precisely, of Chisel --- in the context of Design Space Exploration, and propose a novel design methodology to build custom and adaptable design flows.
    We apply a functional approach to define flexible strategies for design space exploration, based on combinations of basic exploration steps, and provide a proof-of-concept framework along with a library of basic strategies.
    We demonstrate our methodology through several use cases, illustrating how various metrics of interest can be considered to build exploration processes --- in particular, we provide a {\it quality of service}-driven exploration example.

    The methodology presented in this work makes use of designers' expertise to reduce the time required for hardware design, in particular for Design Space Exploration, and its application should ease digital design and enhance hardware developpers' productivity.
\end{abstract}



\maketitle

\section{Introduction}
\label{sec.intro}
    Over the past few decades, software developers have benefited from emerging techniques and semantics, such as oriented programming, functional programming and incremental development.
    In the meantime, the development processes for digital designs have not evolved much, with most still relying on well-known processes based on Hardware Description Languages (HDL).
    To address this problem, and increase design productivity, initiatives have emerged such as Domain Specific Languages (DSL) or High Level Synthesis (HLS).

    The first of these, DSLs, allow users to describe hardware circuits by composing specific operators for a given domain, such as audio, video, or network processing.
    Consequently, this approach offers users some optimized primitives, and the tools implicitly compose those primitives to build the resulting circuit.
    Although this approach is well-suited for users that are less familiar with the task of hardware design, it cannot be adapted for a general use case, as DSL are, by design, restricted to a specific applicative domain.
    Moreover, their use is also constrained by the available primitives (or IPs, for Intellectual Property), as they are often provided by an external vendor.

    In contrast, HLS approaches allow users to define digital designs from a high-level entry point, using languages such as C to algorithmically define how the circuit should behave.
    This type of approach facilitates the design of hardware circuits in most cases, as the implementation details are abstracted from the designers, making it possible for them to focus on functional aspects instead.
    Due to these advantages, in the past decade, HLS tools have grown matured, and they are now widely adopted as an alternative to HDL-based design methodology in the industry. 
    However, HLS approaches are less suitable for some cases, in particular where performance needs are tightly constrained, as the expressivity of the input description --- \ie an imperative description --- is too high-level to specify most of the implementation details.
    Effectively, making those details abstractions can lead to suboptimal designs, especially for domain-specific problems with particular hardware needs, involving for example IP and memory interfacing, tight scheduling or target specific implementations \cite{bruant_towards_2021}.
    Moreover, HLS tools can lack expressivity, as users need to instrument the code to guide the tool toward an acceptable solution, for example by specifying unrolling factors for loops, or data structure partitioning.
    Indeed, as the change of programming paradigm in the flow is a complex problem, it requires complex tool chains, making this instrumentation both tool- and version-dependent.
    As a consequence, adapting a HLS design to a new target (or for other performance needs) can be tedious, with a corresponding impact on the potential for evolution, adaptability and reusability of such solutions.
    In addition, HLS tools have yet to be perfected to fully benefit from state-of-the-art compilation optimizations that could be adapted to the particular context of digital design, including accurate performance models to guide comparisons between implementation candidates \cite{faber_challenges_2022}.

    Based on these concerns, an alternative to both DSL and HLS approaches would appear interesting, to propose a more generic design methodology.
    Such methodology could provide developers with performance, expressivity and reusability, without restriction to a particular applicative domain.
    As an initiative in this direction, Hardware Construction Languages (HCL) have been proposed, which can be used to describe parametrized hardware generators in high-level languages.
    Examples of languages that can be used as HCLs include python, with projects such as MyHDL or PyRTL \cite{jaic2015enhancing, lockhart_pymtl_2014}; Haskell, with C$\lambda$ash \cite{baaij_clash_2010}; and Scala, with SpinalHDL \cite{papon2017spinalhdl} and Chisel \cite{bachrach_chisel_2012}.

    Although these initiatives can effectively be used to add expressivity and reusability to standard design flows, some aspects of the designer's job still need to be considered to propose efficient and generic design methodologies using HCLs.
    Among these aspects, we specifically consider in this work the problem of Design Space Exploration (DSE), which is a key feature in hardware design.
    It consists in selecting, among a (potentially very large) design space composed of functionally equivalent implementations, the implementation which best fits the use case, \ie which has the ``best'' properties (operating frequency, resource usage, latency, etc.) for a particular problem.

    In this work, we propose to consider the use of the HCL paradigm for DSE, to help developers to build and compare equivalent implementations in order to select the best one(s) for a particular use case.
    We chose to work with Chisel, a promising HCL which has imposed itself both in academic and industrial fields, but which does not yet support Design Space Exploration features, as far as we know.
    We start by proposing a new design methodology --- \textbf{meta design} --- which leverages Chisel's features to build circuit generators based on meaningful generation parameters.
    We then use the functional features of the language to propose an alternative to DSE tools.
    We call this process \textbf{meta exploration}.
    It is based on an innovative philosophy: giving more credit to user expertise in guiding the exploration tools, rather than relying on generic heuristics which might not be appropriate for a particular use case.
    This approach allows us to propose a solution that leverages Chisel not only during the design phase, but also to build efficient, expertise-based strategies to explore the possible variations of an architecture.
    We hence propose a dual-methodology for design and exploration, with a particular focus on how a functional approach of the DSE problem can help build concise yet effective design processes.

    In this paper, Section \ref{sec.related} presents an analysis of the Design Space Exploration problem in the literature, while Section \ref{sec.methodology} introduces the meta exploration methodology.
    Section \ref{sec.space} introduces the meta design methodology, the first stage in meta exploration, while Section \ref{sec.dse} introduces the second stage in the proposed methodology, providing a formalization of the DSE problem and how it can be solved through a functional approach.
    Section \ref{sec.qece} introduces {\bf QECE} ({\it Quick Exploration using Chisel Estimators}), the framework that was developed as a proof-of-concept for the meta exploration methodology.
    Finally, Section \ref{sec.expe} demonstrates the usability of the proposed methodology through various experiments, and Section \ref{sec.conclusion} discusses the contributions of this paper, before considering the prospects of the proposed approach.
    
\section{Background and Related Works}
\label{sec.related}
    The process of building a digital circuit to solve a particular use case is central to the role of hardware developers.
    Describing the circuit's micro architecture in a language that can be fed into the design flow is generally relatively easy, but this is only the visible part of a complex process.
    The whole process relies on an in-depth analysis of both the algorithm to be implemented and the targeted technology, guiding the developer's decisions, based on their expertise in digital design to identify the best fit for each use case.

    In this context, the developer is often faced with the task of Design Space Exploration (DSE), which is about comparing and selecting the best implementations for a given use case.
    The implementations compared are selected from among almost equivalent candidates, meaning that the developer needs to generate and compare implementations in a meaningful way.
    However, the problem of exploring this design space is quite broad, and consequently initiatives are being proposed to offer a better comprehension of the problem as a whole.
    As a first approach, Schafer \etal \cite{schafer_high-level_2020} presented the DSE problem as a Multi Objective Optimization Problem (MOOP), with standard objectives to be optimized by the DSE tool.
    They propose an interesting 4-class taxonomy to classify the possible heuristics of exploration: meta heuristics, such as Genetic Algorithms \cite{manuel_model-based_2020, paletti_dovado_2021} or Bayesian optimization \cite{lo_multi-fidelity_2018}, dedicated heuristics \cite{awais_ldax_2021}, supervised learning algorithms \cite{nardi_practical_2019, geng_high-speed_2021}, and graph-based analysis \cite{zhao_performance_2020}.
    This taxonomy highlights the fact that no generic DSE strategy is suitable for every use case, and therefore, to perform efficient DSE, we need generic and parametrizable exploration tools that could be fine-tuned by developers, based on their expertise.

    As the aim of this paper is to propose an alternative to the existing DSE approaches, it is important to define some of the key features that a DSE tool should offer.
    In this context --- and based on the considerations from \cite{schafer_high-level_2020} --- we hence list the characteristics of an ideal DSE tool.
    First of all, it should be \textbf{programmable}, allowing users to parametrize at least two aspects of the exploration: the \textbf{metrics of interest} and the \textbf{exploration strategies}.
    The  metrics of interests --- \ie metrics that the tool should consider during the exploration process --- are key concerns when it comes to exploration, as the user may want to consider standard features of the circuit targeted (\eg resource usage, operating frequency, latency, or power consumption), or more specific features, such as security aspects or quality of service provided.
    The \textbf{exploration strategies} --- \ie the algorithm that the tool uses to scan the design space --- can be used to avoid exhaustive exploration of the design space or suboptimal convergence of the exploration process, for example.
    Furthermore, any such tool should also focus on the \textbf{performance} of the resulting circuits, in order to provide users with exploitable designs.
    A similar focus should also be applied to the \textbf{controllability} of the circuits generated, as the users may want to exploit their expertise to specify some implementation details (\eg memory interface, IP usage or control flow) to guide the tool toward better solutions.
    Last but not least, the DSE tool should be integrable in any development flow, especially in emerging agile approaches, meaning that its results should be readily \textbf{reusable}, \textbf{portable} and \textbf{adaptable} to new use cases (\eg a new technology target, new performance needs, a new functioning environment or even a new tool chain), and should be compatible with most development frameworks.


    As the principle of HLS itself is tightly linked to the problem of exploring design spaces, HLS tools are imposing themselves as turnkey solutions for DSE, with tools produced by both academic \cite{canis2011legup} and industrial stakeholders \cite{zhang_autopilot_2008, singh2011implementing}.
    However, those approaches are limited by design, as inferences from the tool to generate archictecture variations affect both the reusability and the controllability of designs.
    Indeed, HLS tools act on implicit parameters --- known as {\bf exploration knobs} \cite{schafer_high-level_2020} --- to generate several hardware implementations of the same algorithm.
    Exploration knobs --- such as memory partitioning, or the level of unrolling of imperative loops --- can be directly manually tuned by users in the imperative description, but such tuning is highly dependent on the tool and its version, hence influencing the portability of the approach. 
    Moreover, it then requires considerable effort to adapt a description to a new use case, as it is not always straightforward to infer how knob will affect the optimization objective(s) during the exploration process.
    Finally, the available tools are generally based on standard metrics and exploration strategies, that cannot be parametrized by users in a programmatic way.
    Nevertheless, among the DSE initiatives described in the literature, some leverage multiple approaches by combining them for efficient exploration \cite{dong_liu_efficient_2016, bai_boomexplorer_2021}, displaying a need for flexibility in the process of building an exploration strategy.

    Simultaneously, more controllable solutions have emerged, based on hardware-targeting languages rather than higher-level descriptions.
    Among them, Paletti \etal \cite{paletti_dovado_2021} introduced Dovado, an RTL based DSE framework that leveraged HDL parameters for {\bf design space exposition}.
    With Dovado, users can explore a more meaningful design space, and the descriptions can be reused and adapted to new use cases, as they are written in a HDL.
    However, the users have no control on the metrics to be optimized during the exploration process, or on the exploration strategies --- \eg they can use Genetic Algorithm based strategies (based on \cite{shokri_algorithm_2013}), but cannot fine-tune them, or develop and integrate new strategies.
    In addition, the authors claim that they should support a more powerful entry language --- such as Chisel --- as Verilog features are in fact limited for this type of use.

    In this work, to respond to these various needs, we propose a novel approach for a Chisel-based DSE tool focused on user experience, by providing a framework to build flexible, user-defined exploration strategies, based on three main notions.
    First of all, we consider the process of exposing the design space to be explored as a key concern to build efficient exploration strategies --- in contrast to implicit approaches such as HLS where the tools infer the different implementations to be compared, we propose to allow users to define the design spaces themselves, hence relying on their expertise to expose the relevant candidates for exploration.
    Through this method, users can extensively control both the implementation and the design spaces explored by the tool.
    The second notion relates to the comparison of implementation candidates: users should be able to define the metrics to be optimized --- \ie the Objectives of the Multi Objective Optimization Problem \cite{schafer_high-level_2020}.
    Finally, we consider that users should be able to define custom exploration strategies --- \ie user-defined algorithms to scan the design spaces and compare the different implementations --- that could be composed to provide use-case-adapted DSE processes based on the users' knowledge.

\section{Overview of the Proposed Methodology}
\label{sec.methodology}
    The methodology proposed in this paper has two main goals: to build {\bf reusable} and {\bf adaptable} hardware generators, and to use those generators to develop {\bf Design Space Exploration} (DSE) features, making it possible to build flexible design processes.

    \subsection{A Novel Approach to Design Space Exploration}
    \label{sec.methodology:ssec.approach}
    To begin with, we consider the DSE problem from a new perspective, based on the considerations introduced in the previous section --- \ie providing an \textbf{efficient}, \textbf{programmable} and \textbf{reusable} framework for DSE.
    To do, we consider three complementary aspects that can be used to describe an exploration approach:
    \begin{enumerate}
        \item {\bf design space exposition}, which is used to define the architecture variations to be considered in an exploration process.

            In most standard tools, this is usually done through a combination of implicit parameters --- such as the level of loop unrolling in a HLS kernel -- and explicit user guidance --- which can add some hints (usually using {\it pragmas}) to select the best parameters, in order to help the exploration tools to build a meaningful design space.
        \item {\bf metric definition}, which defines the metrics that must be considered to efficiently compare the various implementations.

            Such metrics can be the resource usage --- \eg for developers who want to constrain the area --- the latency of the resulting kernels, or any other metric that makes sense for the specific use case.
        \item {\bf exploration strategy}, which specifies how the exploration tool scans the design space to qualify the different implementations, and how it compares them to identify one or multiple best fit(s).

            As stated by Schafer \etal{} \cite{schafer_high-level_2020}, standard DSE methodologies rely on pre-existing, more or less generic heuristics to propose exploration strategies.
            Such tools do not usually allow users to add a new strategy, or to fine-tune the proposed heuristics for a particular use case.
    \end{enumerate}

    This novel approach to the DSE problem there makes it possible to analyze the existing literature in a new light, as the three aspects are usually considered as a whole.
    In this paper, we leverage this new approach to build a flexible and modular methodology for design space exploration.

    \subsection{Meta Exploration Methodology}
    \label{sec.methodology:ssec.meta-exploration}
\begin{figure}[h!]
        \centering
        \includegraphics[width=0.75\textwidth]{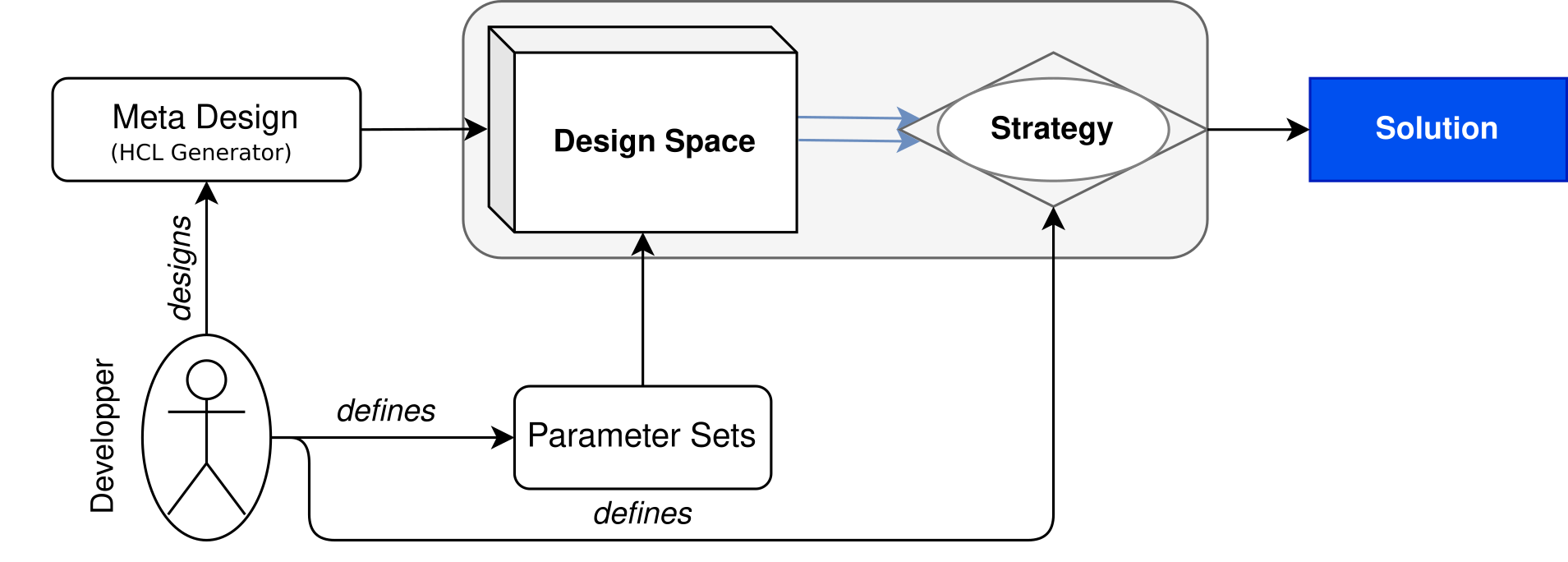}
        \caption{{\bf Meta exploration} methodology}
        \label{sec.methodology:ssec.meta-exploration:fig.meta-exploration}
    \end{figure}

    Based on the considerations set out in Section \ref{sec.methodology:ssec.approach}, we introduce a novel exploration methodology.
    As {\bf Hardware Construction Languages} (HCL) can be used to build hardware generators, and hence expose interesting design spaces to explore, we based this approach on this emerging paradigm.
    This methodology --- that we call {\bf meta exploration methodology} --- is introduced in a simplified schematic in Figure \ref{sec.methodology:ssec.meta-exploration:fig.meta-exploration}, which highlights the process by which the design space for a particular architecture is explored.

    {\bf Meta exploration} involves on two complementary steps, which we will detailed below:
    \begin{enumerate}
        \item exposure of a {\bf design space} that would be interesting to explore, with respect to the algorithm being implemented (Section \ref{sec.space}).
            
            The developer of a module is responsible for providing a parametrized generator along with meaningful parameters, so as to define a design space that only includes meaningful variations of architectures.
        \item leverage of a functional approach to describe an {\bf exploration strategy} as a composition of basic steps (Section \ref{sec.dse}).

            The developer must then define the different steps in the exploration strategy, using a flexible approach to guide and control the exploration process.
    \end{enumerate}

    {\bf Meta exploration} is proposed as a novel methodology which relies on high level features to address the challenges of {\bf Design Space Exploration}.
    The specific goal of this methodology is to allow users to take advantage of their own expertise and knowledge to develop and control the various steps in the design process, rather than relying on more or less configurable steps and inferences.

\section{Design Space Exposition}
\label{sec.space}
    The first step in our novel {\bf meta exploration} methodology is to explicitly define the design space to be explored.
    To perform {\bf design space exposition}, we propose to rely on the developer of the module itself to expose a meaningful design spaces to be explored, by drawing on their experience with respect to the target algorithm, the architectural choices and the target device.

\begin{figure}[h!]
        \centering
        \includegraphics[width=1.0\textwidth]{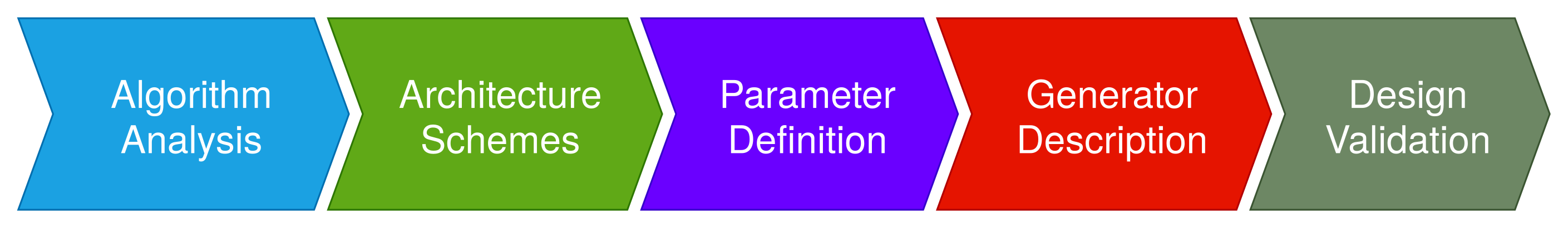}
        \caption{Meta design methodology}
        \label{sec.space:fig.meta-design}
    \end{figure}

    We introduce a sub methodology called {\bf meta design} in Figure \ref{sec.space:fig.meta-design}, which aims to build {\bf highly parametrized} hardware generators based on prior analysis of the algorithm and the functioning environment of the resulting circuit.

    The {\bf meta design} methodology is comparable to standard hardware design processes --- that can be summarized as {\it analysis}, {\it implementation} and {\it validation} (Figure \ref{sec.space:fig.meta-design}) --- as the designer must adequately analyze the target ecosystem before implementing a particular algorithm.
    However, particular emphasis is placed on defining and exposing {\bf high-level parameters} for the generators built, in order to propose architectures variations that can be directly integrated into the generator description.
    Through this approach, it becomes possible to reuse and adapt the circuits and modules produced, hence increasing the designers's productivity.

    This approach mainly makes it possible to perform meaningful {\bf design space exposition} for each module developed, as the parameters are defined by the developers themselves.
    Although this means that the developer must take on a new task --- as exposing relevant parameters can require more reflection and planning than standard RTL design processes --- it also means that any exploration process relying on this methodology will only consider implementations that were selected by the developer of the module(s).

    \begin{figure}[h!]
        \begin{lstlisting}[caption={Example of an annotated Chisel {\bf meta design}},
                           label={sec.space:list.annotations}]
class DummyModule(
    @resource @qos @linear(0, 16) param1: Int,
    @resource      @pow2(0, 8)    param2: Int,
              @qos @enum(4, 6, 9) param3: Int
) extends Module {...}\end{lstlisting}
    \end{figure}

    To outline the relevant parameters as well as the design spaces considered, we introduce an {\bf annotation system} to directly embed the design space in a \textkw{Module} {\bf constructor}.
    In Listing \ref{sec.space:list.annotations}, we present an example of a simple \textkw{Module} with such annotations.
    Three parameters are exposed, and each of them is annotated with some information to guide the exploration:
    \begin{enumerate}
        \item \textkw{param1} is annotated with three indications: the \textkw{@qos} annotation specifies that this parameter affects the {\bf quality of service} of the implementations generated, whereas the \textkw{@resource} annotation indicates that it also influences the resource usage of the different implementations.
            The \textkw{@linear(0, 16)} annotation is used to indicate that \textkw{param1} can take any value within $\llbracket 0, 16\rrbracket $ (\ie 17 possible different values).
        \item \textkw{param2} is annotated with the possible values it can take --- \textkw{@pow2(0, 8)} specifies that its value can be any power of two between $2^0$ and $2^8$ --- \ie 9 possible values.
            It is also annotated with the \textkw{@resource} annotation, meaning that this parameter is expected to have an impact on the resource usage of the implementations generated.
            However, this parameter is not annotated with the \textkw{@qos} annotation, meaning that the developer of the \textkw{DummyModule} does not consider that \textkw{param2} affects the {\bf quality of service} of the implementations generated --- \ie modifying this parameter alone will not change the quality of service of the hardware generated.
        \item \textkw{param3} is also annotated with two indications: it impacts the {\bf quality of service} (using the \textkw{@qos} annotation once again), and it can take any value from among $\{4, 6, 9\}$ --- \ie 3 possible values.
    \end{enumerate}

    These annotations are used to generate the design space to be explored --- in fact, they can even be used to generate multiple design spaces, depending on the exploration concerns that the developer wishes to consider.
    In this example, the user may want to consider all the possible combinations of parameters for the generation of the implementations, or to consider only the \textbf{quality of service} or the \textbf{resource usage} for some particular exploration step, in which case, they only need to consider the parameters that affect those aspects.

    In this example, three different design spaces are considered:
    \begin{itemize}
        \item the ``global'' design space, without consideration of the metrics defined above, which is produced as the {\bf Cartesian product} of all possible values for each parameter, generating all possible combinations of values for the generation parameters.
            For example, in Listing \ref{sec.space:list.annotations}, the ``global'' design space generated would be composed of $17 \times 9 \times 3 = 459$ possible implementations.
        \item the {\bf resource-aware} design space, that only considers the parameters affecting the resource usage of the implementations generated (\ie the parameters annotated with \textkw{@resource}: \textkw{param1} and \textkw{param2}).
            This design space will be composed of fewer possible implementations ($17 \times 9 = 153$), meaning that an exploration strategy focusing only on the resource usage of the implementations explored can benefit from this space reduction to restrict the exploration time.
        \item the {\bf quality of service-aware} design space, that only considers the parameters affecting the quality of service of the implementations generated (\ie the parameters annotated with \textkw{@qos}: \textkw{param1} and \textkw{param3}).
            In this design space, only $17 \times 3 = 51$ implementations are considered.
    \end{itemize}

    This first methodology can therefore be used by developers to perform a meaningful {\bf design space exposition}, based on their own expertise about hardware design and the specific applicative domain targeted.
    This approach allow developers to precisely control the design spaces to be explored, rather than relying on implicit inferences from standard exploration frameworks.

\section{Implementing Meta Exploration using a Functional Approach}
\label{sec.dse}
    The second step of the {\bf meta exploration} methodology is to describe how a design space can be efficiently explored, once again based on user expertise.
    As Chisel is based on Scala, a language which includes functional programming features, we explored the possibilities that this paradigm offers for DSE.
    A functional approach seems particularly appropriate for DSE, if we consider exploration strategies as compositions of functions (\ie mathematical operations) over design spaces. 
    
    In Section \ref{sec.dse:ssec.basis}, we formalize how functional programming can be used to define efficient, user-controlled DSE strategies.
    Section \ref{sec.dse:ssec.func} then provides basic information on functional programming, highlighting how this formalism can be implemented as a programming paradigm to operate over design spaces.
    This formalism is then concretized in Section \ref{sec.dse:ssec.examples}, where we demonstrate how DSE strategies can be built in a powerful, concise and modular way.

    \subsection{Theoretical Basis}
    \label{sec.dse:ssec.basis}
        This section introduces the theoretical basis of this work: it formalizes --- in a mathematical manner --- the key notions required to define an exploration strategy.
        In particular, it defines not only the {\bf exploration strategies}, but also the {\bf design spaces} and the {\bf metrics of interest} as mathematical objects and functions to reason on.


        Let $\mathcal{A}$ be the input vocabulary which will be used to define metric names.
        We define $\mathcal{M}_{\mathcal{A}} = \mathcal{A} \times \mathbb{R}$ the set of {\bf named metrics} with values in $\mathbb{R}$, representing any metric in an exploration process.

        Metrics can be of two kinds: they either refer to the implementation parameters exposed through the {\bf meta design} methodology, or they represent objective and constraint metrics generated during prior exploration steps --- {\bf named metrics} are thus pairs of the form $(name, value)$.

        \example{%
            \textbf{Named metrics} can either be of the form $(param1, 0.0)$ (\textbf{named parameter}\footnotemark) or $(frequency, 247.56)$ (\eg generated from a previous exploration step that estimated the frequency).
        }
            \footnotetext{{\bf Named parameters} are a special case of {\bf named metrics}, as they will represent not only metrics (\ie design properties) but also coordinates in the {\bf design spaces} that are defined in this section.}

        \addtocounter{footnote}{-1}
        Let $n \in \mathbb{N^*}$.
        We define a {\bf configuration of order n} --- \ie a configuration relying on {\bf n named parameters}\footnotemark --- as $x_n = \{x_0, ..., x_{n-1}\}$, with $x_i \in \mathcal{M}_{\mathcal{A}}$ and $i \in \llbracket 0, n-1\rrbracket $.
        Each configuration stands for a distinct implementation variation, and we thus define the design space as corresponding to all possible implementations for a given {\bf meta design}.

        \example{%
            In this module, there are \textbf{3 named parameters}, thus we consider \textbf{configurations of order 3}.%
            \newline{}
            A possible configuration in the design space is $\{(param1, 0.0), (param2, 64.0), (param3, 6.0)\}$. 
        }

        We then define a {\bf point of order (n, k)} as representing an improved configuration, bearing both the configuration parameters $x_i$ with $i \in \llbracket 0, n-1\rrbracket $ and some generated metrics $m_i$ with $i \in \llbracket 0, k-1\rrbracket $.
        A point $p_{(n, k)}$ can then be defined as a vector of elements in $\mathcal{M}_{\mathcal{A}}$ which characterizes a given implementation, as exposed in Equation \ref{sec.dse:ssec.basis:eq.point}. 

        \definition{
            \begin{equation}
                \label{sec.dse:ssec.basis:eq.point}
                p_{(n, k)} = \{\underbrace{x_0, ..., x_{n-1}}_{n\: \text{parameters}}, \underbrace{m_0, ..., m_{k-1}}_{k\: \text{metrics}}\}
            \end{equation}
        }
        
        \example{%
            After estimating both the percentage of used LUTs and the operating frequency, a point from the exposed design space could be:
            \begin{equation*}
                p_{(3, 2)} = \{\underbrace{(param1, 0.0), (param2, 64.0), (param3, 6.0)}_{3\: \textrm{parameters}}, \underbrace{(freq, 247.56), (\%LUT, 0.77)}_{2\: \textrm{metrics}}\}
            \end{equation*}
        }

        Based on this definition, we characterize a {\bf design space $s_n$ of order n} as being a {\bf set of points of order n} (Eq. \ref{sec.dse:ssec.basis:eq.space}).
        The number of {\bf points} in a {\bf design space} is given by $\#(s_n)$.

        \definition{
            \begin{equation}
                \label{sec.dse:ssec.basis:eq.space}
                s_n = \{p_{(n, \_)}\}
            \end{equation}
        }

        \example{%
            This definition is sufficiently powerful to express the three design spaces that were constructed in Section \ref{sec.space}:
            \begin{itemize}
                \item the global design space $s_{global}$ of order \textbf{3}, with $\#(s_{global}) = 17 \times 9 \times 3 = 459$:
                    
                    \begin{equation*}
                        s_{global} = \{\{\underbrace{(p1, 0.0), (p2, 1.0), (p3, 4.0)}_{\textrm{3 parameters}}, \underbrace{\dots}_{\textrm{k metrics}}\}, \dots, \{\underbrace{(p1, 16.0), (p2, 64.0), (p3, 9.0)}_{\textrm{3 parameters}}, \underbrace{\dots}_{\textrm{k metrics}}\}\}
                    \end{equation*}
                \item the \textbf{resource aware} design space $s_{resource}$ of order \textbf{2}, with $\#(s_{resource}) = 17 \times 9 = 153$:
                    
                    \begin{equation*}
                        s_{resource} = \{\{\underbrace{(p1, 0.0), (p2, 1.0)}_{\textrm{2 parameters}}, \underbrace{(p3, 4.0), \dots}_{\textrm{k metrics}}\}, \dots, \{\underbrace{(p1, 16.0), (p2, 64.0)}_{\textrm{2 parameters}}, \underbrace{(p3, 4.0), \dots}_{\textrm{k metrics}}\}\}
                    \end{equation*}
                \item the \textbf{quality of service aware} design space $s_{qos}$ of order \textbf{2}, with $\#(s_{qos}) = 17 \times 3 = 51$:
                    
                    \begin{equation*}
                        s_{qos} = \{\{\underbrace{(p1, 0.0), (p3, 4.0)}_{\textrm{2 parameters}}, \underbrace{(p2, 1.0), \dots}_{\textrm{k metrics}}\}, \dots, \{\underbrace{(p1, 16.0), (p3, 9.0)}_{\textrm{2 parameters}}, \underbrace{(p2, 1.0), \dots}_{\textrm{k metrics}}\}\}
                    \end{equation*}
            \end{itemize}
            }

        We only consider the number of parameters for each configuration when defining the dimensions (\ie order) of a design space, as the metrics do not represent dimensions but only information on the designs.
        For generalization purposes, we define $\mathbb{S}_n$ as the set of all possible spaces $s_n$.

        We now wish to define {\bf design space exploration strategies} operating on design spaces defined in this way.
        We start by defining {\bf cost functions $c$} as a way to generate new {\bf named metrics} in $\mathcal{M}_{\mathcal{A}}$.
        As shown in Equation \ref{sec.dse:ssec.basis:eq.cost}, {\bf cost functions} take points in a design space --- \ie a list of {\bf $k+n$ named metrics}\footnote{where \textbf{k} is the number of \textbf{named parameters}, \ie the parameters used to generate the \textbf{meta circuit}, and \textbf{m} is the number of \textbf{named metrics} that are not parameters} --- and use them to compute a new {\bf named metric}.

        \definition{
            \begin{equation}
                \label{sec.dse:ssec.basis:eq.cost}
                \begin{split}
            c: {\mathcal{M}_{\mathcal{A}}}^{n+k} & \rightarrow {\mathcal{M}_{\mathcal{A}}}\\
            p_{(n,k)} & \mapsto c(p_{(n,k)})
                \end{split}
            \end{equation}
        }

        \example{%
            An example of \textbf{cost function} could be used to compute the \textit{efficiency} of a \texttt{DummyModule} implementation.
            To do so, the $c_{eff}$ function could be built using previously computed metrics such as a \textit{frequency estimation} (denoted by $p_{freq}$, \ie the metrics named $freq$ in a point $p$) or \textit{resource usage} ($p_{\%LUTS}$).

            \begin{equation*}
                c_{eff} = \frac{p_{freq}}{p_{\%LUTs}}
            \end{equation*}
        }

        Using such cost functions, we define {\bf estimation transforms of order $\theta$}, which are used to enhance a given design space with {\bf $\theta$ new metrics}.
        Given {\bf $\theta$ cost functions} $c_i$ with $i \in \llbracket 0, \theta-1\rrbracket $, an estimation transform $f_\theta$ operating over points $p_{(n, k)}$ is defined in Equation \ref{sec.dse:ssec.basis:eq.transform}.

        \definition{
            \begin{equation}
                \label{sec.dse:ssec.basis:eq.transform}
                \begin{split}
                    f_\theta: {\mathcal{M}_{\mathcal{A}}}^{n+k} & \rightarrow {\mathcal{M}_{\mathcal{A}}}^{n+k+\theta}\\
                    p_{(n, k)} & \mapsto p_{(n, k + \theta)}
                \end{split}
            \end{equation}
        }

        The resulting points are thus enhanced with $\theta$ new {\bf named metrics}, as shown in Equation \ref{sec.dse:ssec.basis:eq.enhanced}, \ie they now bear not only their {\bf generation parameters}, but also $k + \theta$ {\bf named metrics}.

        \definition{
            \begin{equation}
                \label{sec.dse:ssec.basis:eq.enhanced}
                \begin{split}
                    p_{(n, k + \theta)} &= \{x_0, ..., x_{n-1}, m_0, ..., m_{k-1}, c_0(p_{(n, k)}), ..., c_{\theta-1}(p_{(n, k)})\}\\
                                       &= \{\underbrace{x_0, ..., x_{n-1}}_{n\: \text{parameters}}, \underbrace{m_0, ..., m_{k-1}}_{k\: \text{old metrics}}, \underbrace{m_k, ..., m_{k+\theta-1}}_{\theta\: \text{new metrics}}\}
                \end{split}
            \end{equation}
        }

        \example{%
            An \textbf{estimation transform} $f_{eff}$ of \textbf{order 1} can use the cost function $c_{eff}$ to enhance every point in a design space $s_n = \{p_{(n, k)}\}$, by adding an \textit{efficiency} metric.
    
            \begin{align*}
                \textrm{Let } p & = \{\underbrace{(p1, 0.0), (p2, 1.0), (p3, 4.0)}_{\textrm{3 parameters}}, \underbrace{(freq, 247.56), (\%LUTs, 0.77)}_{\textrm{2 metrics}}\}\\
                \textrm{and } c_{eff}(p) & = \frac{p_{freq}}{p_{\%LUTs}} = \frac{247.56}{0.77} = 321.50\\
                f_{eff}(p) & = \{\underbrace{(p1, 0.0), (p2, 1.0), (p3, 4.0)}_{\textrm{3 parameters}}, \underbrace{(freq, 247.56), (\%LUTs, 0.77), (eff, 321.50)}_{\textrm{3 metrics}}\}\\
            \end{align*}
        }

        In Equation \ref{sec.dse:ssec.basis:eq.morphism}, we define a {\bf morphism of order n} as a modification of a {\bf design space of order $n$}.
        A morphism can be used to sort, prune or even enhance the design space, meaning that we do not impose any hypothesis on the cardinality $\#(s'_{n'})$ of the resulting {\bf design space} with respect to the original cardinality of the space $\#(s_n)$.
        In addition, a {\bf morphism} can also modify the dimensions of the {\bf design space}, hence changing the number of {\bf dimensions} $n'$ to be explored.
        We call $\mathbb{M}_n$ the set of all possible {\bf morphisms} of order n.

        \definition{
            \begin{align}
                \label{sec.dse:ssec.basis:eq.morphism}
                \begin{split}
                    m_n: \mathbb{S}_n & \rightarrow \mathbb{S}_{n'}\\
                    s_n & \mapsto s_{n'}'
                \end{split}
            \end{align}
        }

        \example{%
            An example of morphism in the \texttt{DummyModule} induced design space can be the construction of the \textbf{resource-aware} design space from the global one (\textbf{order 3}):
            \begin{align*}
                m_{resource} : & \; \mathbb{S}_3 && \to \mathbb{S}_2\\
                & \;s_{global} && \mapsto s_{resource}\\
                & \{\underbrace{(p1, 0.0), (p2, 1.0), (p3, 4.0)}_{\textrm{3 parameters}}, \underbrace{\dots}_{\textrm{k metrics}}\} && \mapsto  \{\underbrace{(p1, 0.0), (p2, 1.0)}_{\textrm{2 parameters}}, \underbrace{(p3, 4.0), \dots}_{\textrm{k +1 metrics}}\}
            \end{align*}

            It is important to note that the morphism $m_{resource}$ reduces the dimensions of the design spaces by removing one dimension (\ie one parameter).
            Another example of morphism of order 3, but that does not modify the dimensions of the design spaces, could be a simple pruning of the points $p$ for which the estimated frequency $p_{freq}$ is lower than 200 MHz:
            \begin{align*}
                m_{pruning} : & \; \mathbb{S}_3 \to \mathbb{S}_3\\
                & p \mapsto \begin{cases}
                    \;p \textrm{ if } p1 >= 200.0\\
                    \;\emptyset \textrm{ otherwise }
                \end{cases}
            \end{align*}
        }
        
        We finally define how the {\bf estimation transforms} should be applied over a design space, before making any potential modification through a {\bf given morphism}.
        Considering an estimation transform $f_\theta$ of order $\theta$, a morphism $\mu_n$ of order n, and an input {\bf design space} $s_n$ of order n, we define a {\bf transform application function} $a_{(n, \theta)}$ of order $(n, \theta)$ in Equation \ref{sec.dse:ssec.basis:eq.application}.

        \definition{
            \begin{align}
                \label{sec.dse:ssec.basis:eq.application}
                \begin{split}
                    a_{(n, \theta)}: \mathbb{F}_\theta \times \mathbb{M}_n \times \mathbb{S}_n & \rightarrow \mathbb{S}_{n'}\\
                    (f_\theta, \mu_n, s_n) & \mapsto \mu_n(\{f_\theta(p_{(n, k)})\}) \: \text{with} \: p_{(n,k)} \in s_n  
                \end{split}
            \end{align}
        }
        
        \example{%
            We can thus compose the previously built cost function $c_{eff}$ and the morphism $m_{pruning}$ to compute the efficiency of every point in our design space $s_{global}$ with an estimated frequency exceeding 200 MHz:

            \begin{align*}
                a_{eff\geq200} : &\; \mathbb{S}_3 &&\to \mathbb{S}_3 \\
                &\; s_{global} &&\mapsto m_{pruning}(\{f_{eff}(p) \in s_{global}\})
            \end{align*}
        }

        It is important to note that no assumptions can be made about how the morphism and the estimation transform are applied over the input design space, and how they will interact.
        For example, the estimation transform can be applied to all the points in the design space before modifying its structure, or it can be applied through a more selective approach, for example using a gradient descent algorithm.
        Hereafter, we will denote the set of all the possible {\bf transform application functions} of order $(n, \theta)$ as $\mathbb{A}_{(n, \theta)}$.

        To keep the formalism concise, we will use the {\bf currying} notion, which is used in the {\bf functional programming paradigm}.
        It refers to the action of converting a function with multiple arguments to a set of parametrized functions, which only take one argument.
        For example, a function $f(a, b)$ can be converted to a set of functions $f_a$, which can then be applied to the second argument, $b$, meaning that $f(a, b) = f_a(b)$.

        We will thus convert our {\bf transform application functions} into simple functions operating over an input {\bf design space}, as shown in Equation \ref{sec.dse:ssec.basis:eq.app-currified}.

        \definition{
            \begin{equation}
                \label{sec.dse:ssec.basis:eq.app-currified}
                \begin{split}
                    a_{(n, \theta)}(f_\theta, \mu_n, s_n) \Rightarrow a_{(n, \theta)}(f_\theta, \mu_n)(s_n) = \alpha_{(f_\theta, \mu_n)}(s_n)
                \end{split}
            \end{equation}
        }

        With this overall formalization, we can now represent every possible mathematical operation over a design space that may be needed to define an exploration strategy.

        Using all these constructs, we define an {\bf exploration step} as a {\bf function} operating over a {\bf design space} by applying, given some {\bf transform application functions}, a set of {\bf estimation transforms} to the {\bf points} making up the space, before modifying its structure by applying a given {\bf morphism}, and producing a new space enhanced with {\bf new metrics}.
        Equation \ref{sec.dse:ssec.basis:eq.step} formalizes the notion of {\bf exploration step} with respect to the theoretical bases set out in this section.

        \definition{
            \begin{equation}
                \label{sec.dse:ssec.basis:eq.step}
                \begin{split}
                    e: \mathbb{M}_{n'} \times \mathbb{A}_{(n, \theta)} \times \mathbb{S}_n & \rightarrow \mathbb{S}_{n''}\\
                    (m_{n'}, \alpha_{(f_\theta, \mu_n)}, s_n) & \mapsto m_{n'}(\alpha_{(f_\theta, \mu_n)}(s_n))
                \end{split}
            \end{equation}
        }
        
        \example{%
            The \textbf{transform application function} $a_{eff\geq200}$ can itself be considered as an exploration strategy, returning the implementations that have a sufficient \textit{operating frequency}, along with their \textit{efficiency}.
            However, it can also be composed with an other exploration strategy, \eg to sort the resulting design space according to the \textit{efficiency}, using a morphism $m_{sort}$.

            \begin{align*}
                m_{sort} : & \; \mathbb{S}_3 \to \mathbb{S}_3\\
                &  \; s \mapsto s'
            \end{align*}
            {\flushright where $s'$ is an ordered version of $s$, with respect to the efficiency of each implementation ($p_{eff}$)}

            We can then define an exploration strategy $e_{eff,\; sorted}$:
            \begin{align*}
                e_{eff,\; sorted} :&\; \mathbb{S}_3 \to \mathbb{S}_3\\
                &\;s \to m_{sort} \circ m_{pruning} (f_{eff}(\{p \in s\})
            \end{align*}

            In this example, the initial design space $s$ is first pruned using the $m_{prune}$ morphism, removing all the implementations that operate under 200MHz, before the $m_{sort}$ morphism is applied to sort the resulting design space based on the \textit{efficiency} of the remaining solutions.
        }

        In Equation \ref{sec.dse:ssec.basis:eq.step-currified}, we use {\bf currying} once again to define {\bf exploration steps} as simple functions operating over {\bf design spaces}.

        \definition{
            \begin{equation}
                \label{sec.dse:ssec.basis:eq.step-currified}
                \begin{split}
                    e(m_{n'}, \alpha_{(f_\theta, \mu_n)}, s_n) \Rightarrow \epsilon_{(m_{n'}, \alpha)}(s_n)
                \end{split}
            \end{equation}
        }

        We finally use {\bf functional programming} to compose basic strategies and build more complex ones, by applying $n$ exploration strategies $\llbracket \epsilon_0, ..., \epsilon_{n-1} \rrbracket $ sequentially over an initial {\bf design space}.

    \subsection{Bases of Functional Programming for Design Space Exploration}
    \label{sec.dse:ssec.func}
        To make the most of the functional programming paradigm to solve the DSE problem, we provide some basic functions for a concise description of some popular programming patterns.
        For each function introduced, we will propose multiple equivalent descriptions --- which are more or less compact and understandable --- to help the user understand how this emerging paradigm can be used for DSE.

        First of all, we consider the {\bf map-reduce} pattern, where a function is applied to every element in a given sequence, before performing a reduction to return only one value.
        For example, considering a vector of elements $e_i$, $i \in \llbracket 0, k\rrbracket $, this pattern can be used to compute a sum of squares, as in Equation \ref{sec.dse:ssec.func:eq.usual}.%
        \footnote{In this context, we will consider a simplification that is often used in functional programming, to replace implicit parameters (\eg x) by a simple placeholder \_, when there is no ambiguity for the compiler.}

        \begin{equation}
            \label{sec.dse:ssec.func:eq.usual}
            \begin{split}
                sum &= e.map(x \Rightarrow x^2).reduce((a, b) \Rightarrow a+b)\\
                    &= e.map(\_^2).reduce(\_+\_)\\
                    &= e.map(square).reduce(add)\\
                    \text{with}\: &square(x) = x^2 \: \text{and} \: add(a, b) = a+b
            \end{split}
        \end{equation}

        We will also use some simple operations that can be applied to various collections, for example the {\tt sortWith} function, which operates over a collection {\tt col} to sort its elements by applying a comparison function.
        For example, if we want to sort a collection {\tt col} of objects using a particular attribute {\tt .value}, the patterns described in Equation \ref{sec.dse:ssec.func:eq.sortWith} can be used.

        \begin{equation}
            \label{sec.dse:ssec.func:eq.sortWith}
            \begin{split}
                newCol &= col.sortWith((a, b) \Rightarrow a.value \leq b.value)\\
                       &= col.sortWith(\_.value \leq \_.value)\\
                       &= col.sortWith(compare)\\
                   \text{with}\: &compare(a, b) = a.value \leq b.value
            \end{split}
        \end{equation}

        Another useful operation is the possibility to filter a collection (Eq. \ref{sec.dse:ssec.func:eq.filter}), using a boolean function --- \eg to select only the elements for which the {\tt .value} attribute is above a threshold $min_{value}$.

        \begin{equation}
            \label{sec.dse:ssec.func:eq.filter}
            \begin{split}
                newCol &= col.filter(x \Rightarrow x.value > min_{value})\\
                       &= col.filter(\_.value > min_{value})\\
                       &= col.filter(func)\\
                   \text{with}\: &func(x) = x.value > min_{value}
            \end{split}
        \end{equation}

        With respect to the formalism introduced in the previous section, {\tt sortWith}, {\tt map} and {\tt filter} can all be defined as {\bf morphisms}, if the collection {\tt col} is a design space.
        As those constructs do not modify the number of {\bf parameters} in the points they are operating on, they can even be considered {\bf endomorphisms} --- \ie morphisms from $\mathbb{S}_n$ to $\mathbb{S}_n$.

        In the following section, we will use a compact description of the various functions to be applied to the design spaces, to demonstrate how the {\bf functional programming paradigm} can help users to define concise yet intelligible exploration strategies.

    \subsection{Application examples: Building Complex Strategies using Functional Programming}
    \label{sec.dse:ssec.examples}

        As an example of application of this programming model, we define {\bf exhaustive strategies} in Equation \ref{sec.dse:ssec.examples:eq.exhaustive}, where the {\bf transform application function} (from Eq. \ref{sec.dse:ssec.basis:eq.application}) consists in an exhaustive application (\ie \texttt{map}) of the {\bf estimation transform} $f_\theta$ to all the {\bf points} in the {\bf design space} $s_n$.

        \begin{equation}
            \label{sec.dse:ssec.examples:eq.exhaustive}
            exhaustive_{(m_n, f_\theta)}(s_n) = m_n(s_n.map(f_\theta)))
        \end{equation}

        The morphism $m_n$ makes post-processing (\eg sorting or pruning) of the design space possible after the application of $f_\theta$.
        It can be applied to define how to {\bf exhaustively sort} a {\bf design space $s_n$}, based on a comparison function $cmp$ used to define a {\bf custom order} over $s_n$ (Eq. \ref{sec.dse:ssec.examples:eq.sort}).

        \begin{equation}
            \label{sec.dse:ssec.examples:eq.sort}
            \begin{split}
                sort_{(f_\theta, cmp)}(s_n) &= exhaustive_{(sortWith(cmp), f_\theta)}(s_n)\\
                                          &= s_n.map(f_k).sortWith(cmp)
            \end{split}
        \end{equation}
        
        The {\bf exhaustive pruning} of a space --- \ie boolean partitioning of a design space --- is defined in Equation \ref{sec.dse:ssec.examples:eq.prune}.
        This definition is based on a pruning function $f_{prune}$, used as a filtering criterion to specify which {\bf points} should be left in the resulting design space.

        \begin{equation}
            \label{sec.dse:ssec.examples:eq.prune}
            \begin{split}
                prune_{(f_\theta, f_{prune})}(s_n) &= exhaustive_{(filter(f_{prune}), f_\theta)}(s_n)\\
                                                   &= s_n.map(f_\theta).filter(f_{prune})
            \end{split}
        \end{equation}

        Based on these two basic strategies, we can define a more complex one, which uses both quick metric generation through {\bf Register-Transfer Level} (RTL) estimations of the resources, and accurate estimations through synthesis processes.

        To do so, we define a first strategy $\epsilon_0$ which we will refer to as the {\it preliminary pruning}, and a second one, $\epsilon_1$, that will be termed as the {\it refinement}.
        We define $estim$ as a {\bf cost function} of order 1, based on an RTL estimation of DSP resource usage, producing a metric named $DSP_{estim}$ for a given circuit.
        We also define $synth$ as another {\bf cost function} of order 1, that calls an external synthesis tool to produce a metric named $DSP_{synth}$, which is the reference value that the DSP estimation should approximate.
        
        $\epsilon_0$ is defined in Equation \ref{sec.dse:ssec.examples:eq.e0}: considering a threshold $DSP_{max}$ which represents the maximum amount of DSP acceptable in an implementation, this exploration strategy aims to prune the design space to remove every implementation that is estimated to use too many DSPs.

        \begin{equation}
            \label{sec.dse:ssec.examples:eq.e0}
            \begin{split}
                \epsilon_0(s_n) & = prune_{(estim, \_.DSP_{estim} < DSP_{max})}(s_n) \\
                                & = s_n.map(estim).filter(\_.DSP_{estim} < DSP_{max})
            \end{split}
        \end{equation}

        $\epsilon_1$ is defined in Equation \ref{sec.dse:ssec.examples:eq.e1}.
        It is used to compare and sort all the implementations in a design space, with respect to the {\it real} DSP usage --- \ie the DSP usage as estimated by a synthesis tool, which should be more accurate than the RTL-based estimation  {\it estim}.
        This exploration strategy can then be used to select the implementation using the least DSP resources in a design space.

        \begin{equation}
            \label{sec.dse:ssec.examples:eq.e1}
            \begin{split}
                \epsilon_1(s_n) & = sort_{(synth, \_.DSP_{synth} > \_.DSP_{synt})}(s_n) \\
                                & = s_n.map(synth).sortWith(\_.DSP_{synth} > \_.DSP_{synth})
            \end{split}
        \end{equation}
        
        Consequently, a global {\bf exploration strategy} $\epsilon_\tau = \epsilon_0 \circ \epsilon_1$ can be defined as a composition of $\epsilon_0$ and $\epsilon_1$ in Equation \ref{sec.dse:ssec.examples:eq.etau}.
        This strategy will initially prune the design space of the implementations that are too ``large'' to fit into the available DSPs, using a rapid, RTL-based estimation of the required resources.
        Subsequently, it will synthetize all the remaining implementations in the design space, and sort this space to select the ``smallest'' implementation.

        \begin{equation}
            \label{sec.dse:ssec.examples:eq.etau}
            \begin{split}
                \epsilon_\tau(s_n) = s_n&.map(estim).filter(\_.DSP_{estim} < DSP_{max})\\
                                        &.map(synth).sortWith(\_.DSP_{synth} > \_.DSP_{synth})
            \end{split}
        \end{equation}

    This methodology makes it possible to describe and compose {\bf exploration strategies} in a functional way, as each can be considered as a simple function applied to a{\bf design space}.
    Moreover, each individual strategy can be defined in a functional way, as it is mainly defined as a combination of various {\bf estimation transforms} and some {\bf morphisms} applied to a given {\bf design space}.
    Hence, this formalism highlights the path to build a flexible and modular DSE framework, where the user can fine-tuned and compose each step, based on their experience.

\section{Demonstrator Framework}
\label{sec.qece}
    To demonstrate the usability of the {\bf meta exploration} methodology, we introduce {\bf QECE} ({\it Quick Exploration using Chisel Estimators}).
    QECE is a Chisel-based framework designed to allow users to concisely build custom and flexible DSE strategies, based on the steps presented in Sections \ref{sec.space} and \ref{sec.dse}.
    It is distributed as an {\bf open-source} Scala package \cite{ferres_qece_2021}, that can easily be imported into any Chisel-based project.

    As described in Section \ref{sec.dse}, we consider two notions to be the keys to define an efficient and adaptable exploration process: the {\bf metrics of interest}, and the {\bf exploration strategy}.
    Indeed, an efficient exploration strategy should rely on {\bf user-defined metrics} and {\bf estimators}, as the developer of a module is the best placed to specify which properties are of interest for exploration, and to indicate how they should be estimated, depending on the accuracy and performance required.
    Moreover, the person charged with the exploration should also be able to apply their expertise with respect to the module being explored to {\bf guide the exploration process}, by providing the algorithm to scan the design space structure and apply the various estimators.
    
    To meet these two requirements, QECE mainly relies on two complementary {\bf built-in libraries}, allowing users to build exploration strategies in line with their use cases:
    \begin{enumerate}
        \item a library of {\bf estimators}, that can be used to estimate user-defined {\bf metric(s) of interest} (Section \ref{sec.qece:ssec.estimators})
        \item a library of {\bf exploration steps}, that can be composed in a functional way to build complex strategies, using the formalism introduced in Section \ref{sec.dse} (Section \ref{sec.qece:ssec.strategies})
    \end{enumerate}

    \subsection{Library of estimators}
    \label{sec.qece:ssec.estimators}
        The {\bf metric estimators} provided with QECE are introduced in Table \ref{sec.qece:ssec.estimators:table.estimators}.
        Three different abstraction levels are considered to integrate these estimators:
        \begin{itemize}
            \item {\bf FIRRTL level} --- \ie operations on FIRRTL representations \cite{izraelevitz_2017_reusability} (the intermediate representation used by Chisel)
            \item {\bf Simulation level} --- \ie empirical estimations based on the analysis of simulation results
            \item {\bf Register Transfer Level} --- \ie any process operating on an RTL (Verilog/VHDL) description.
        \end{itemize}

        \begin{table}[h!]
            \centering
            \begin{tabular}{|c|c|c|}
                \hline
                \ccg {\bf Abstraction level} & \ccg {\bf Estimation method} & \ccg {\bf Metric estimated} \\
                \hline
                \multirow{2}*{FIRRTL} & IR analysis & Resource usage \\
                ~ & Analytical formulas & Custom metrics \\
                \hline
                Simulation & Empirical approach & Quality of service \\
                \hline
                \multirow{2}*{Register Transfer Level} & \multirow{2}*{Synthesis} & Resource usage \\
                ~ & ~ & Operating frequency \\
                \hline
            \end{tabular}
            \caption{Built-in estimators in QECE}
            \label{sec.qece:ssec.estimators:table.estimators}
        \end{table}

        These estimators are provided as a {\it proof-of-concept} library, to demonstrate the applicability of the approach to basic use cases.
        However, the framework is designed to be easily \textbf{extendable}, providing a flexible Application Programming Interface for users to develop and integrate their own estimation methods --- which should operate at one of these three levels of abstraction --- for {\bf particular use cases} considering {\bf specific metrics}.

    \subsection{Library of exploration steps}
    \label{sec.qece:ssec.strategies}

        For modularity purposes, we also propose a library of {\bf basic exploration steps} that are integrated in QECE.
        Each step is briefly introduced in Table \ref{sec.qece:ssec.strategies:table.strategies}, and their use is further detailed below.
        
        \begin{table}[h!]
            \centering
            \begin{tabular}{|c|c|c|}
                \hline
                \ccg {\bf Exploration strategy} & \ccg {\bf Description} & \ccg {\bf Functional equivalent} \\
                \hline
                Exhaustive mapping & Apply a function to the whole design space & {\tt map} \\
                \hline
                Exhaustive sorting & Sort the whole design space & {\tt sort} \\
                \hline
                Exhaustive pruning & Partition the whole design space & {\tt filter} \\
                \hline
                \multirow{2}*{Gradient sort} & Space sorting operation based on & \multirow{2}*{-} \\
                ~ & a gradient approach & ~ \\
                \hline
                \multirow{2}*{Quick pruning} & Space pruning operation based on & \multirow{2}*{-} \\
                ~ & a gradient approach & ~ \\
                \hline
            \end{tabular}
            \caption{Built-in exploration strategies in QECE}
            \label{sec.qece:ssec.strategies:table.strategies}
        \end{table}

        The three first steps --- {\it exhaustive mapping}, {\it sorting} and {\it pruning} --- correspond to commonly used functional constructs that were introduced in Section \ref{sec.dse:ssec.func}.
        These constructs are used to provide the basic operations to {\bf exhaustively cover} the collections used --- \ie the design spaces.

        Two more complex steps are also provided to demonstrate how the functional approach considered can be used to build adaptable exploration strategies.
        As those steps involve more advanced {\bf topology considerations} over the design spaces, we start by defining two main functions to be used by the exploration algorithms:\footnote{QECE API also enables user to specify their own topologies for design spaces, as building a new exploration strategy could require to change the data structure underneath the design space for performance purposes.} 
        \begin{enumerate}
            \item $S.getNeighbours(p: Point, n: Norm, d: Int)$, which defines the neighborhood of a point $p$ in a particular design space $S$.

                The neighborhood is defined with respect to a norm $n$ --- \eg $\|.\|_1$ for the Manhattan distance, or $\|.\|_\infty$ for the Chebyshev distance --- and a maximal distance from the point $p$.
            \item $isOnFrontier(p: Point)$, which defines wether a point $p$ in a design space is on the frontier that partitions this space in two.
                The two partitions respectively include points that are not pruned by the filtering action, and those that are.

                Considering a Boolean function $filter$ which states wether a point $p$ should be pruned from the design space or not, a point $p$ is on the frontier if it satisfies Equation \ref{sec.qece:ssec.strategies:eq.isOnFrontier}.
                This equation states that $p$ is on the frontier if it is not pruned ($\neg filter(p)$) and if at least one of its neighbors $q$ (with respect to the Chebyshev distance $\|.\|$) is pruned (\ie $filter(q)$). 
        \end{enumerate}

        \begin{equation}
            \label{sec.qece:ssec.strategies:eq.isOnFrontier}
            isOnFrontier(p) \iff \neg filter(p) \land \exists q \in S.getNeighbours(p, \|.\|_\infty, 1) / filter(q)
        \end{equation}

        The {\it gradient sort} strategy applies a gradient descent to a design space, with the aim of finding a local optimum for the cost function being optimized.
        This strategy can be used to sort a design space more quickly than by an exhaustive approach, in particular when the estimations to be performed during an exploration step are complex and time consuming.
        It is important to note that even if this approach searches for a local optimum, all the implementations that are estimated as part of the process are sorted and returned in the result, thus maximizing the information made available to users.

        The algorithm used for this type of sorting is introduced in Algorithm \ref{sec.qece:ssec.strategies:alg.gradient}.
        The \textsc{Gradient} procedure (line 1) requires an initial design space $S$, a cost function $f$ to compare the different implementations, and an optional starting point $x$.
        It returns a new design space $S'$ , which is sorted with respect to $f$.
        The first step (lines 2-4) is the initialization, where the resulting space $S'$ is an empty set and the cost of the starting point ($x$ if provided, or the first point in the design space $S$) is computed and stored as the current optimum.
        The main computations then occur in a loop (lines 6-20), which is guaranteed to terminate as the space is finite --- meaning that, in the worst case, every point is considered before $S'$ is returned (line 18).
        In this loop, the cost of every neighbor of the current optimum is computed (line 9) and compared (line 10) in order to select a new optimum.
        If a new optimum is found among the neighbors of the current optimum (line 13), the current optimum is updated (line 15) and the gradient descent continues.
        However, if no new optimum is found, it means that the current optimum is also a local optimum, and the gradient descent stops (line 18).
        The resulting design space $S'$ is then returned, composed of all the points for which the cost was estimated during the process (line 11).

        \noindent\underline{NB:} The actual implementation relies on a local cache to optimize computation time, and also exhibits a {\bf parallelism} parameter to leverage multi-threading and take full advantage of the available resources while comparing the neighbors of the current optimum.

        \begin{algorithm}[ht!]
            \caption{Gradient descent algorithm}
            \label{sec.qece:ssec.strategies:alg.gradient}
            \begin{minipage}{1.0\linewidth}
                \Input \\
                \Desc{S}{design space to explore}\\
                \Desc{f}{cost function to sort S}\\
                \Desc{\{x\}}{optional starting point for the descent}\\
                \Output\\
                \Desc{S'}{sorted (and pruned) design space}
                \begin{algorithmic}[1]
                    \PROCEDURE{Gradient}{$S: Space, f: Point \Rightarrow Double, \{x: Point\}$}
                        \STATE $S' \leftarrow \emptyset$ \MyComment{the result set is empty at first}
                        \MyLineComment{either use x as starting point, or the head of S}
                        \STATE $(current, cost) \leftarrow x~?~(x, f(x)) : (S[0], f(S[0]))$
                        \MyLineComment{iterate until a local optimum is found}
                        \WHILE{$True$} 
                            \MyLineComment{the space is finite; an optimum exists}
                            \STATE $neighbours \leftarrow S.getNeighbours(current, \|.\|_1, 1)$
                            \STATE $costs \leftarrow neighbours.map(f)$ \MyComment{apply f to all neighbors}
                            \STATE $index \leftarrow indexWhere(costs.max)$ \MyComment{select best neighbor}
                            \STATE $S' \leftarrow S' + neighbours$
                            \MyLineComment{a neighbor is better than the current implem.}
                            \IF{$costs[index] > cost$} 
                                \MyLineComment{update current and cost with best neighbor}
                                \STATE $(current, cost) \leftarrow (neighbours[index], costs[index])$
                            \ELSE
                                \MyLineComment{return sorted resulting space with respect to $f$}
                                \RETURN $S'.sort$ 
                            \ENDIF
                        \ENDWHILE
                    \ENDPROCEDURE
                \end{algorithmic}
            \end{minipage}
        \end{algorithm}

        The {\it quick pruning} strategy is the most complex strategy included in this library, demonstrating how specific a strategy can be if needed.
        Like the {\it gradient sort}, this strategy relies on neighborhood exploration to partition a design space into two compact parts, using a Boolean function to define wether an implementation should be pruned or not.
        \begin{algorithm}[ht!]
            \caption{Quick pruning algorithm}
            \label{sec.qece:ssec.strategies:alg.quick}
            \begin{minipage}{1.0\linewidth}
                \Input \\
                \Desc{S}{design space to explore} \\
                \Desc{f}{pruning function to discriminate space} \\
                \Output \\
                \Desc{S'}{pruned design space}
                \begin{algorithmic}[1]
                    \PROCEDURE{QuickPruning}{$S: Space, f: Point \Rightarrow Boolean$}
                        \MyLineComment{try to find a starting point on the frontier}
                        \PROCEDURE{Start}{}
                            \MyLineComment{explore a sub space to find the starting point (a diagonal between extrema)}
                            \STATE $diag \leftarrow S.getDiagonal(S.min, S.max)$
                            \MyLineComment{select the first non-pruned point on the diag.}
                            \STATE $p \leftarrow diag.filter(!f)[0]$
                            \MyLineComment{if a frontier exists, it crosses this diag. either directly, or in the neighborhood}
                            \IF{$isOnFrontier(p)$}
                                \RETURN $p$
                            \ELSE
                                \RETURN $S.getNeighbours(p, \|.\|_\infty, 1).filter(!f)[0]$
                            \ENDIF
                        \ENDPROCEDURE
                        \MyLineComment{iteratively build the frontier}
                        \PROCEDURE{Frontier}{$p: Point$}
                            \STATE $(currents, frontier) \leftarrow ([p], [p])$
                            \WHILE{$!currents.isEmpty$}
                                \MyLineComment{explore neighborhoods to find frontier points}
                                \STATE $n \leftarrow currents.flatMap(S.getNeighbours(\_, \|.\|_\infty, 1)).filter(!f)$
                                \STATE $onFrontier \leftarrow n.filter(isOnFrontier) - frontier$
                                \STATE $frontier \leftarrow onFrontier + frontier$
                                \MyLineComment{update with the new limits of the frontier}
                                \STATE $currents \leftarrow onFrontier$
                            \ENDWHILE
                            \RETURN $frontier$ \MyComment{a frontier has been found}
                        \ENDPROCEDURE
                        \PROCEDURE{Update}{$frontier: [Point]$}
                            \MyLineComment{select only points above the frontier}
                            \RETURN $S.filter(isAbove(frontier))$
                        \ENDPROCEDURE
                        \RETURN \textsc{Update}(\textsc{Frontier}(\textsc{Start}))
                    \ENDPROCEDURE
                \end{algorithmic}
            \end{minipage}
        \end{algorithm}

        The algorithm --- introduced in Algorithm \ref{sec.qece:ssec.strategies:alg.quick} --- is based on strong hypotheses about the design space being explored, but can be used to more rapidly partition a design space than the exhaustive approach.
        It is similar to the Pareto approximation approach proposed by Ye \etal{} \cite{ye_scalehls_2021}, which iteratively uses space sampling to find some Pareto optimal points before exploring their neighborhoods to approximate the frontier.

        The main procedure, \textsc{QuickPruning}, requires an initial design space $S$ and a pruning function $f$ to produce a new design space $S'$, which should include only points that are not pruned (\ie points $p$ for which $f(p) = False$).
        This procedure is based on three sub-procedures: \textsc{Start} (lines 3-14), \textsc{Frontier} (lines 16-25), and \textsc{Update} (lines 28-31).

        Assuming that a single frontier separates pruned and non-pruned implementations in the given space, it can be pruned by applying the pruning function to a fraction of the points, leading to a faster convergence.
        To do so, the first step is to identify a first point on the frontier, which is done by the \textsc{Start} procedure (lines 3-14), using a simple assumption: if a single frontier exists, then the frontier crosses the diagonal subspace composed of points ranging from the minimal to the maximal configuration (with respect to the implementation {\bf parameters}).
        This procedure therefore scans this diagonal subspace to apply the filtering criterion $f$ and select the first point on the diagonal which is not pruned (line 7).
        This first point is either on the frontier that we want to build (lines 9-10), or one of its neighbors is on the frontier (lines 11-12), meaning that we have found a point on the frontier.
        After identifying this first point, we iteratively build the frontier using the \textsc{Frontier} procedure, which uses neighborhood exploration to build it step by step (lines 18-25), as we assume that the frontier is continuous --- as part of the required hypothesis.
        To do so, each step of the loop considers all the neighbors of all the points on the current frontier, and checks wether those neighbors are on the frontier (line 20).
        The new current frontier is hence built by adding the neighbors that were also found to be on the frontier (\ie points $p$ for which $isOnFrontier(p) = True$) to the previous current frontier (line 22).
        This step is repeated until an iteration identifies no new points (line 18), meaning that no point in the neighborhood of the current frontier is on the frontier we are building.
        The last step updates the design space (using the \textsc{Update} procedure, lines 28-31), to retain only the points that are ``above'' the computed frontier.

    \subsection{Complex strategy building}
    \label{sec.qece:ssec.build}

\begin{figure}[h!]
            \begin{subfigure}{1.0\textwidth}
                \includegraphics[height=0.2\textheight]{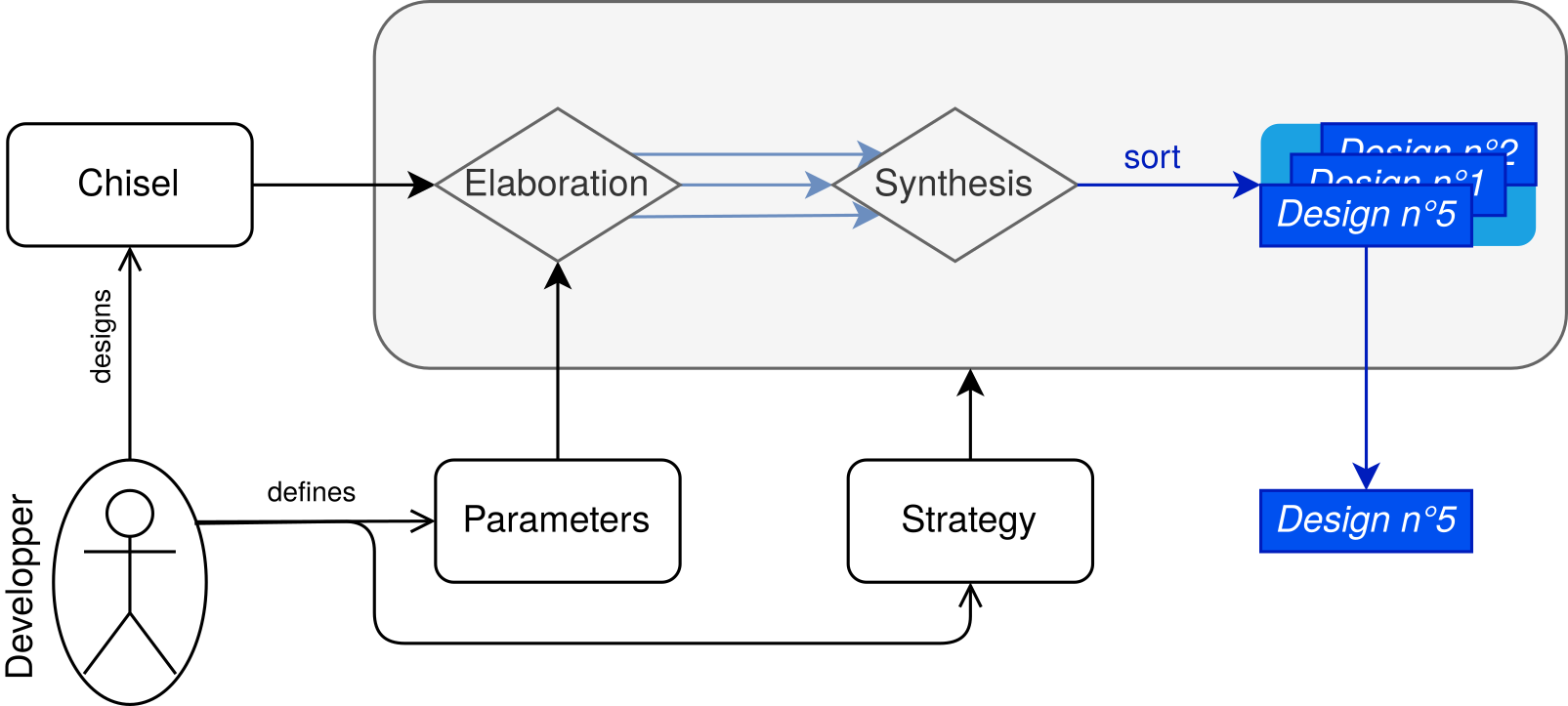}
                \caption{Simple exhaustive strategy ($\epsilon_1$ --- Eq. \ref{sec.dse:ssec.examples:eq.e1})\vspace{0.3cm}}
                \label{ssec.qece:ssec.build:fig.complex:sfig.exhaustive}
            \end{subfigure}
            \begin{subfigure}{1.0\textwidth}
                \includegraphics[height=0.2\textheight]{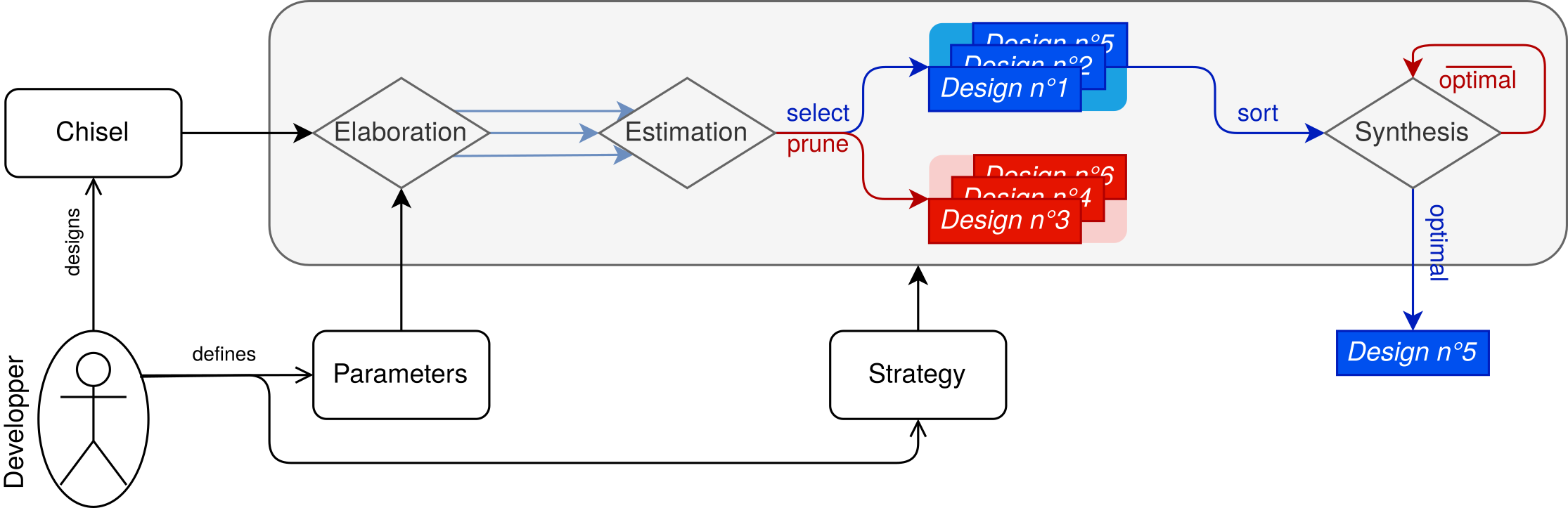}
                \caption{Gradient descent based strategy}
                \label{sec.qece:ssec.build:fig.complex:sfig.gradient}
            \end{subfigure}
            \caption{Example of QECE-powered complex strategies}
            \label{sec.qece:ssec.build:fig.complex}
        \end{figure}

        Based on these two libraries --- one for {\bf estimation purposes}, the other for {\bf exploration strategies} --- the developers of a Chisel module can now easily build an exploration strategy meeting their requirements for a particular use case.
        To demonstrate this feature, Figure \ref{sec.qece:ssec.build:fig.complex} introduces two examples of {\bf meta exploration strategies} that can be built using QECE.

        The first strategy (Figure \ref{ssec.qece:ssec.build:fig.complex:sfig.exhaustive}) is a naive approach to exploration, as it relies on synthesizing each possible implementation to compare the results and select the best (for the function that the developers wish to optimize).
        It could actually be an implementation of the $\epsilon_1$ strategy (Eq. \ref{sec.dse:ssec.examples:eq.e1}), for an exploration process where the developer is looking for the implementation that consumes the fewest DSP blocks possible. 
        This approach is guaranteed to select the best implementation in the design space, but will require long exploration processes, as the synthesis processes are costly to perform on large designs.
        
        In constrast, the second strategy (Figure \ref{sec.qece:ssec.build:fig.complex:sfig.gradient}) is more complex, in order to reduce the number of synthesis processes to run and thus speeds-up the exploration processes.
        It relies on three sequential steps, demonstrating the possibilities and the concision of the proposed approach:
        \begin{enumerate}
            \item the resource usage of each implementation is estimated at the {\bf FIRRTL level} (as defined in Table \ref{sec.qece:ssec.estimators:table.estimators}).
                Designs that do not fit onto the target device board are pruned, without running long synthesis processes on the whole design space. 
                This step is actually an implementation of the {\it preliminary pruning} step $\epsilon_0$, as defined in Eq. \ref{sec.dse:ssec.examples:eq.e0}.
                However, it considers not only DSP usage, but also the usage of the other resources available on the target device.
            \item the resulting space --- pruned of overly large designs --- is sorted to select the largest implementation remaining in the design space, and present it as the first element of the space
            \item a {\bf gradient-based approach} is used to find a local optimum for throughput, using the synthesis results to compute the resource usage and operating frequency.
                Through this approach, fewer implementations have to be actually synthesized to identify the optimum.
                Moreover, as described in Algo. \ref{sec.qece:ssec.strategies:alg.gradient}, this strategy uses the first element of the space --- which was sorted in the previous step --- as the starting point for the gradient descent, which further speeds-up the exploration process.
        \end{enumerate}

        This strategy is particularly suitable for circuit generators where some of the resources available on the target device are critical.
        For example, if we consider the implementation of accelerators relying on multipliers, we can assume that the FPGA logic synthesis flow will map those operators to {\bf Digital Signal Processing} (DSP) blocks.
        In fact, those resources can usually be exploited to compute multiplications faster than using {\bf Look-Up Tables} (LUT).
        Furthermore, they can be pipelined by the synthesis tool to reduce the delay path.
        We can use this assumption on resource usage to identify implementations that would consume too many {\bf DSP} blocks, but also to sort the remaining implementations to prioritize those that use many {\bf DSP} blocks. 
        This approach therefore makes it possible to prune implementations that are too large to fit on the target device, and to avoid considering suboptimal implementations that do not use all the available DSP blocks, as the gradient approach will not consider them.

        These examples demonstrate the methodology introduced in Section \ref{sec.dse}, showing how basic steps can be composed to define more complex strategies, that meet the needs of the actual exploration use case.
        However, as these examples are simple, we introduce multiple use cases in the next section, along with QECE-based solution to efficiently solve them.

\section{Experiments and Results}
\label{sec.expe}
    In this section, we introduce a number of use cases to demonstrate application of QECE in realistic scenarios.
    As part of this demonstration, to help developers build and compare their own strategies, we also provide a Chisel-based {\bf open-source benchmark} \cite{ferres_benchmark_2021}, including several kernel generators that were built by applying the {\bf meta design} methodology (Section \ref{sec.space}):
    \begin{itemize}
        \item Black Scholes computations
        \item Fast Fourier Transform algorithm
        \item General Matrix Multiply algorithm
    \end{itemize}

    Section \ref{sec.expe:ssec.black-scholes} introduces a {\bf quality of service}-aware exploration use case on a {\bf Black Scholes} computation kernel, based on a {\bf Monte Carlo} generic implementation.
    In addition to this use case, Section \ref{sec.expe:ssec.others} compares a number of exploration strategies on other concerns and kernels.

    \subsection{Building an Expertise-based Exploration Strategy with QECE: a Monte Carlo Powered Use Case}
    \label{sec.expe:ssec.black-scholes}

        In this section, we detail the application of the {\bf meta exploration} methodology to a specific use case, to help the reader to understand the whole flow for a QECE-based exploration.

        For this experiment, exploration processes were run on a {\bf 24-cores} (48-threads) server running at {\bf 3.2 GHz}, with {\bf 188 GB} of RAM.
        A {\bf 2-hour timeout} was applied for synthesis processes to avoid memory crashes due to non-converging syntheses, as memory usage itself is not limited.

        \begin{figure}[h!]
            \centering
            \includegraphics[height=0.24\textheight]{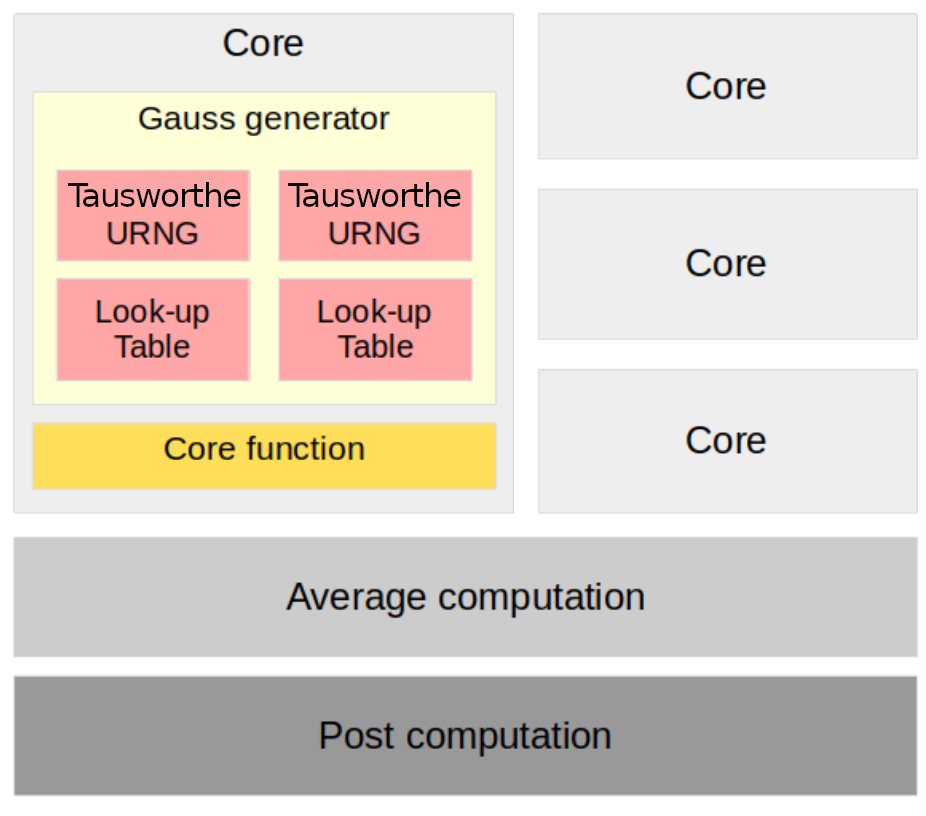}
            \caption{Simplified architecture for the Monte Carlo generator}
            \label{sec.expe:ssec.black-scholes:fig.archi}
        \end{figure}

        We consider a {\bf Monte Carlo}-based {\bf Black Scholes} computation kernel generator implemented in Chisel.
        The aim of the kernels is to accelerate computation of the {\bf Black Scholes} formula, used to estimate the theoretical value of an option, given some applicative parameters (Eq. \ref{sec.expe:ssec.black-scholes:eq.std}).\footnote{The values $\mu$, $\sigma$ and $T$ are parameters of the Black-Scholes model itself, and are considered constants in this implementation.}
        \begin{equation}
            \label{sec.expe:ssec.black-scholes:eq.std}
            S(t) = S(0)\times e^{(\mu - \frac{1}{2}\sigma^2)T + \sigma\sqrt{T}\mathcal{N}(0, 1)}
        \end{equation} 
        
        However, as hardware-based computations of the exponential function are costly, we leverage the {\bf Euler-Maruyama} method \cite{hu_semi_1996} to iteratively approximate the formula (Eq. \ref{sec.expe:ssec.black-scholes:eq.euler}).
        The number of iterations of this method can be seen as a {\bf generation parameter} in this use case, affecting both the {\bf latency} and the {\bf accuracy} of estimations.
        \begin{equation}
            \label{sec.expe:ssec.black-scholes:eq.euler}
            S_{\Delta t} = S_0((1 + (\mu - \frac{1}{2}\sigma^2)\Delta t) + \sigma\sqrt{\Delta t}\mathcal{N}(0, 1)
        \end{equation}

        The proposed {\bf meta design} is based on the architecture schematic presented in Figure \ref{sec.expe:ssec.black-scholes:fig.archi}, using a fixed number of {\bf computation cores} running in parallel, each relying on a {\it Pseudo Random Number Generator} (PRNG) to estimate the target value.
        As the estimation formulas rely on {\bf normal distributions}, a {\bf Gauss generator} is proposed.
        For efficient generation, each \textbf{PRNG} is based on {\bf Tausworthe sequences} to generate {\bf uniform distributions} \cite{tezuka_1991_tausworthe}, which are then fed into pre-computed {\bf Look-Up Tables} to generate {\bf normal distributions}, using the {\bf Box Muller method} \cite{box_1958_note}.
        Each {\bf core} is therefore composed of a {\bf PRNG} and a {\bf core function}, which are used to estimate a statistical value --- with respect to the core configuration.
        Once a fixed number of estimations has been produced by the parallel cores, they are used to compute an {\bf average value} (using the {\bf average computation unit}) --- that can then be {\bf post-processed} to improve estimation quality, through the {\bf post-computation unit}.
        In this kernel, multiple parameters can be tuned to adjust the estimation method, and hence approximate the target value within a user-defined error range.
        
        \begin{figure}[h!]
            \begin{lstlisting}[xleftmargin=0mm,
                               caption={Expertise-based design space for Black Scholes {\bf meta design}},
                               label={sec.expe:ssec.black-scholes:list.space}]
class BlackScholesTopLevel(
    @resource @qos @linear(8, 32)  dynamic: Int,
    @resource @qos @linear(8, 32)  precision: Int,
              @qos @pow2(5, 10)    nbIteration: Int,
              @qos @pow2(1, 6)     nbEuler: Int,
    @resource      @pow2(2, 10)    nbCore: Int
) extends Module {...} \end{lstlisting}
        \end{figure}
        \vspace{-0.6cm}

        Using the proposed architecture, we expose an explorable design space for the {\bf Black Scholes meta design} (see Section \ref{sec.space}), which is based on {\bf 5 parameters}, as introduced in Listing \ref{sec.expe:ssec.black-scholes:list.space} and discussed below.
        Both the {\it dynamic} and the {\it precision} parameters are used to define the {\bf data representation}, using a {\bf signed fixed point} data type.
        The {\it nbIteration} parameter defines the total number of estimations used to compute a single value, whereas the {\it nbCore} sets the number of parallel cores to be used for estimations.
        Finally, the {\it nbEuler} parameter defines the number of inner iterations to be perform on each core, to approximate the value of the exponential function (Eq. \ref{sec.expe:ssec.black-scholes:eq.euler}).

        In this use case, we consider two different concerns that will be used for exploration: the {\bf quality of service} --- \ie the error rate induced by both the Monte Carlo and the Euler Maruyama methods --- and the {\bf resource usage} --- \ie the use of the available resources on the target FPGA board.
        These two concerns are directly indicated in the design space, using the annotations \textkw{@resource} and \textkw{@qos} on the parameters --- a parameter annotated with \textkw{@resource} is considered by the developer as affecting {\bf resource usage} metrics, as explained in Section \ref{sec.space}.

        \definecolor{qeceKeyword}{RGB}{255,0,0}
        \definecolor{customKeyword}{RGB}{255,0,0}
        \begin{figure}[h!]
            \begin{lstlisting}[xleftmargin=0mm,
                               basewidth={0.55em, 0.5em},
                               firstnumber=0,
                               numbersep=2.5mm,%
                               numberstyle=\small,%
                               numbers=left,%
                               caption={Expertise-based strategy to explore Black Scholes implementations},
                               label={sec.expe:ssec.black-scholes:list.strategy}]
val builder  = <@\textcolor{qeceKeyword}{StrategyBuilder()}@>
val strategy = builder.<@\textcolor{qeceKeyword}{buildStrategy}@>(
  builder.<@\textcolor{qeceKeyword}{quickPrune}@>[BlackScholesTopLevel](
    <@\textcolor{customKeyword}{QualityOfService.simulation}@>,
    _.error > 0.05,
    metric = Some(new qos)          // relative to @qos annotation
  ),
  builder.<@\textcolor{qeceKeyword}{reduceDimension}@>[BlackScholesTopLevel](new resource, true),
  builder.<@\textcolor{qeceKeyword}{map}@>[BlackScholesTopLevel](<@\textcolor{customKeyword}{Transforms.latency}@>), 
  builder.<@\textcolor{qeceKeyword}{sort}@>[BlackScholesTopLevel](
  <@\textcolor{qeceKeyword}{TransformSeq.empty}@>,
    m => m("dynamic") + m("precision") + m("nbCore"),
    (_ < _)
  ),
  builder.<@\textcolor{qeceKeyword}{gradient}@>[BlackScholesTopLevel](
    <@\textcolor{qeceKeyword}{TransformSeq.synthesis}@> ++ <@\textcolor{customKeyword}{Transforms.throughput}@>
    func = _("throughput"),
    cmp = (_ > _)
  )
)\end{lstlisting}
        \end{figure}

        After defining the design space using the {\bf meta design} methodology, we showcase QECE's ability to build an ad hoc exploration strategy adapted to a particular use case, in Listing \ref{sec.expe:ssec.black-scholes:list.strategy}.\footnote{Red keywords are objects, classes and methods that are provided by QECE libraries.} %
        This strategy is based on {\bf 5 sequential steps}, which have been defined based on our expertise and knowledge on both the algorithm and the target board:
        \begin{enumerate}
            \item \underline{\bf lines 2-6:} the design space is pruned to remove implementations that are estimated to induce an error of more than {\bf 5\%} (line 4).

                For this pruning step, we use the {\bf quick pruning} algorithm (Algo. \ref{sec.qece:ssec.strategies:alg.quick}) to rapidly partition the space (line 2).
                Moreover, as we only consider the quality of service of the different implementations, we specify that the framework can eliminate all the dimensions that are not annotated with \textkw{@qos} (line 5).
                We use a custom-defined \textkw{QualityOfService.simulation} transform to compute the quality of service (line 3).
            \item \underline{\bf line 7:} this line specifies how the dimensions of the design space are reduced for the remaining exploration steps.
                All the {\bf meta design parameters} (see Figure \ref{sec.expe:ssec.black-scholes:list.space}) not annotated with \textkw{@resource} have no effect on the resource metrics, and are thus removed from the dimensions. 
                They are in fact the parameters acting on the number of iterations: {\it nbIteration} and {\it nbEuler}.
                At this point, we can state two things: all the remaining designs are acceptable with respect to the required quality of service for this use case, and reducing the number of iterations can only improve both resource usage and throughput.%
                \footnote{Increasing the number of iterations can only improve the quality of service at the cost of increased {\bf latency}, but we already have a sufficiently low error after the pruning step.}

                The Boolean parameter in the \textkw{context.reduceDimension} method specifies that the dimension removal will project the parameters onto the {\bf minimal values} in the space --- as we wish to keep the number of iterations as low as possible among the remaining designs.

                Removing those two dimensions from the design space means that the number of remaining implementations is considerably reduced.
            \item \underline{\bf line 8:} we compute the latency of each remaining implementation.
            \item \underline{\bf lines 9-13:} we select the ``minimal point'' of the remaining design space (\ie the point with the minimal sum of parameters) and place it in front of the design space, to use it as a starting point in the next step.

                At this step in the exploration process, we consider that the resource usage will grow with the parameters, and we want to keep it as low as possible --- hence minimizing the remaining parameters is a straightforward approach to accelerate convergence of the next step.
            \item \underline{\bf lines 14-18:} we use a {\bf gradient descent} algorithm (Algo. \ref{sec.qece:ssec.strategies:alg.gradient}) to identify a local optimum with respect to the objective of the exploration, \ie optimizing the throughput of the circuits generated while minimizing resource usage --- in continuation of the previous step, where the design space was sorted to minimize resource usage.

                For this final step, syntheses are required to provide a realistic estimation of both resource usage and operating frequency, and it is thus necessary to adopt a clever strategy to limit the number of costly process runs.
                We therefore use neighborhood explorations to iteratively identify an acceptable solution for the use case, and return it to the users.
        \end{enumerate}

        \begin{table}[h!]
            \scalebox{1.0}{
                \begin{tabular}{|c|ccc|cc|c|}
                    \hline
                    \ccg ~ & \ccg ~ & \ccg ~ & \ccg {\bf Throughput} & \multicolumn{2}{c}{\ccg {\bf Area}} & \ccg ~ \\
                    \multirow{-2}*{\ccg {\bf Rank}} & \multirow{-2}*{\ccg {\bf Parameters}} & \multirow{-2}*{\ccg {\bf Error}} & \ccg ($est.s^{-1}$) & \ccg Max \% & \ccg Resource & \multirow{-2}*{\ccg {\bf Frequency}} \\
                    \hline
                    1 & [12, 21, 64, 2, 64] & 5.34\% & 125.06 & 25\% & DSP & 250.13 MHz \\
                    2 & [12, 20, 64, 2, 64] & 4.6\% & 125.06 & 25\% & DSP & 250.13 MHz \\
                    3 & [12, 22, 64, 2, 64] & 6.54\% & 125.03 & 25\% & DSP & 250.06 MHz \\
                    4 & [13, 21, 64, 2, 64] & 6.06\% & 125.03 & 25\% & DSP & 250.06 MHz \\
                    5 & [12, 22, 64, 2, 32] & 5.34\% & 62.53 & 12.5\% & DSP & 250.13 MHz \\
                    \hline
                \end{tabular}
            }
            \caption{Best implementations found using the exploration strategy from Listing \ref{sec.expe:ssec.black-scholes:list.strategy}.}
            \label{sec.expe:ssec.black-scholes:table.results}
        \end{table}

        Table \ref{sec.expe:ssec.black-scholes:table.results} presents the top 5 implementations with respect to our use case objectives.
        As can be seen, the result of an exploration is a data frame summarizing the different metrics estimated during the exploration process.
        This frame can be pruned and/or sorted to provide the users with an understandable overview of the results.

        In this particular example, the reader will observe that three out of the five best implementations introduce an error of more than {\bf 5\%}, even though one of the exploration goals was to remain under this threshold. 
        This outcome is due to the approximations made by the {\bf quick pruning algorithm} (Algo. \ref{sec.qece:ssec.strategies:alg.quick}), which can result in a imprecise pruning.
        However, this approximation leads to a small error overhead, while accelerating the exploration process, and it is up to the designer to indicate whether this level of overhead is acceptable in their particular use case.

        Moreover, we can also note that the top 4 implementations only use 25\% of the available resources, even though one could expect to improve the throughput by replicating computation units until the resources are saturated.
        This is due to the {\bf meta architecture} of the Monte Carlo kernel itself, which limits the number of parallel cores in the kernel to the number of inner estimations needed before computing the average value --- as this number of estimations only affects the quality of the results, it is fixed (here at 64) once the exploration process has identified kernels satisfying the {\bf quality of service} requirements.
        As a consequence, the exploration tool cannot consider Monte Carlo kernels with more than 64 cores --- however, it also means that the designer can use the results of this exploration process to conclude that they need to replicate the whole kernel 4 times to saturate the available resources and optimize the throughput.

        Through this use case, we demonstrate the concision of the functional approach to DSE, showing how QECE can be used to describe a complex, user-based strategy in a few lines, while providing meaningful results to the designers.

    \subsection{Exploring Multiple Kernels through Various Considerations}
    \label{sec.expe:ssec.others}
        For the following experiments, the exploration processes were run on a server including {\bf 6 cores} (8 threads) running at {\bf 3.46 GHz} and {\bf 78 GB} of RAM.
        As in the previous experiment, a {\bf 2-hour timeout} was applied for synthesis processes.

        \vspace{-0.2cm}
        \begin{table}[ht!]
            \centering
            \scalebox{1.0}{
                \centering
                \begin{tabular}{|c|c|c|c|c|c|c|c|}
                    \hline
                    \ccg ~ & \ccg ~ & \ccg ~ & \ccg {\bf \#[space]} & \ccg {\bf \#synth} & \ccg ~ & \ccg ~ \\
                    \multirow{-2}*{\ccg \bf Kernel} & \multirow{-2}*{\ccg \bf Strategy} & \multirow{-2}*{\ccg \bf Best throughput} & \ccg (\#dimension) & \ccg (\#timeout) & \multirow{-2}*{\ccg \bf Time} & \multirow{-2}*{\ccg \bf Speed-up} \\
                    \hline
                    \multirow{2}*{FFT128} & {\it Exhaustive} & {\it 1.767 Tb/s} & \multirow{2}*{7 (1)} & {\it 7 (0)} & {\it 00h22m45s} & - \\
                    ~ & Gradient & 1.767 Tb/s & ~ & 3 (0) & 00h19m51s & $\times$1.15 \\
                    \hline
                    \multirow{2}*{FFT512} & {\it Exhaustive} & {\it 5.479 Tb/s} & \multirow{2}*{9 (1)} & {\it 9 (0)} & {\it 02h11m51s} & - \\
                    ~ & Gradient & 5.479 Tb/s & ~ & 3 (0) & 02h18m29s & \inred{$\times$0.95} \\
                    \hline
                    \multirow{2}*{GEMM} & {\it Exhaustive} & {\it 231.334 GOp/s} & \multirow{2}*{41 (2)} & {\it 41 (19)} & {\it 13h51m56s} & - \\
                    ~ & Gradient & 231.334 GOp/s & ~ & 6 (1) & 03h21m06s & $\times$4.1 \\
                    \hline
                \end{tabular}
                }
                \captionsetup{justification=centering}
                \caption{Comparing various exploration strategies without quality of service concerns.\\{\it (Exhaustive strategies are used as baselines)}}
                \label{sec.expe:ssec.others:table.explo}
        \end{table}
        \vspace{-0.8cm}

        Table \ref{sec.expe:ssec.others:table.explo} compares the exploration strategies introduced in Figure \ref{sec.qece:ssec.build:fig.complex} on two different kernel generators, namely the {\bf Fast Fourier Transform} (FFT) generator and the {\bf General Matrix Multiply} (GEMM) generator.\footnote{The results of the GEMM-based explorations were already published in a previous work \cite{ferres_2021_integrating}.}

        To begin with, the {\bf FFT meta design} relies on a size parameter --- \eg FFT128 and FFT512 --- and we do not provide a metric that can be used to compare two implementations of different sizes.
        Consequently, two different design spaces must be explored, each exposing a small number of different implementations to be considered in the processes.
        More importantly, for a given FFT size, the exploration processes only consider {\bf one parameter} when generating the design spaces\footnote{The only parameter considered for generation is actually the kernel I/O bandwidth.}, meaning that we can consider them as mono dimensional explorations.
        As can be seen from the data, applying our custom exploration strategy to those design spaces leads to the same optimal implementation (in terms of {\bf throughput}) than with the exhaustive approach, but with almost no gain (or even a small overhead) in terms of process duration.

        In constract, the second experiment considers the exploration of a {\bf GEMM} generator.
        In this case, the design space is actually a bit wider, with 41 different implementations to deal with and compare, in two dimensions, as two distinct parameters are considered\footnote{We consider both the size of the input matrices and the kernel I/O bandwidth as generation parameters.}.
        In this context, applying our custom exploration strategy results in $\times$7 less synthesis than the exhaustive approach, providing a $\times$4 speed-up factor, while finding the same best implementation in terms of throughput.

        These experiments reveal that a key feature in efficient exploration processes is exploration of an interesting design space, relying on the users' expertise to provide explorable {\bf meta designs}.
        If this is properly done, our results also demonstrate that we can rely on QECE's concision to provide efficient exploration strategies, improving designers' productivity by reducing the time spent on design space exploration.

    \subsection{Synthesis on the Experiments}
    \label{sec.expe:ssec.synthesis}
        Our results demonstrate the usability of QECE, our Chisel-based design space exploration framework, in 3 different use cases, using custom designed hardware generators.
        In these use cases, we show how user expertise can be leveraged to develop efficient exploration processes that can be fine-tuned by the designers of a kernel.
        More specifically, users can specify the {\bf metrics} they wish to optimize --- \eg resource usage, operating frequency, throughput --- and how they want to optimize them, by describing the {\bf exploration strategies}.
        
        We provide integrated libraries for both aspects in QECE, and demonstrate how those basic components can be composed to create complex, user-defined exploration strategies that can be used to efficiently explore a given design space.
        We also highlight the advantage for kernel designers to expose interesting design spaces, in order to guide the exploration tool toward a best fit in an acceptable amount of time.

        Moreover, we demonstrate the concision of our proposal, highlighting how a functional approach to the design space exploration problem can help to describe custom exploration strategies in a concise way.
        This approach is valuable in terms of reusability, as it allows reuse and adaptation of an exploration strategy for a new use case.

        The experiments presented demonstrate the modularity, flexibility and concision of the proposed approach, through various use cases and exploration concerns.

\section{Conclusion and Perspectives}
\label{sec.conclusion}
    In this work, we propose a new approach --- called {\bf meta exploration} --- to solve the problem of Design Space Exploration (DSE), based on the use of Chisel, a Hardware Construction Language (HCL).
    Specifically, we consider the use of functional programming, a powerful programming paradigm well known to software developers, in the context of hardware description.
    With meta exploration, we allow users to draw on their own expertise to describe custom and adapted exploration strategies, in contrast to most recent exploration tools, which rely on more or less implicit heuristics for DSE.

    By considering a user-centered approach, we aim to improve the designers' productivity by reducing exploration times, allowing users to precisely describe the aspects they wish to optimize, how they want to do it, and how the exploration processes should interact with the hardware they design.
    We believe that this approach, combined with the use of a recent Hardware Construction Language --- which can considerably improve the reusability of the circuits produced --- will make hardware development easier.

    To prove our claims, we provide {\it Quick Exploration using Chisel Estimators} (QECE), a Chisel-based framework which implements our functional approach to DSE, and demonstrate its use in several cases.
    Our results show that this framework can be used to speed-up exploration processes by leveraging specific knowledge on the algorithms implemented, to build efficient exploration strategies from realistic scenarios.
    Moreover, we demonstrate that specific metrics, such as \textit{quality of service}, can be considered in the exploration processes, highlighting the flexibility of our approach.

    The framework proposed in this work is a proof of concept, but it is built in a modular and extensible way.
    As the literature behind the DSE problem is extensive, QECE could therefore greatly benefit from users implementing existing heuristics, algorithms and metrics.
    For this reason, we provide QECE as an open-source solution \cite{ferres_qece_2021} along with an applicative benchmark \cite{ferres_benchmark_2021}, and we strongly encourage readers to experiment using the framework on their own projects, and to contribute to its further development.
    
\clearpage
\bibliographystyle{ACM-Reference-Format}
\bibliography{article}

\end{document}